\documentclass[journal]{IEEEtran}

\usepackage{amsmath,graphicx,stmaryrd}
\usepackage{amssymb}
\usepackage{times}
\usepackage{graphicx}
\usepackage{subfig}
\usepackage{tabularx}
\usepackage{multirow}
\usepackage{textcomp}
\usepackage{setspace}
\usepackage{url}
\usepackage{color}
\usepackage{tikz}
\usepackage{pgfplots}

\newcolumntype{?}{!{\vrule width 1pt}}

\definecolor{color1}{rgb}{0,0,1}
\definecolor{color2}{rgb}{1,0,0}
\definecolor{color3}{rgb}{1,0,1}
\definecolor{color4}{rgb}{0,1,1}
\definecolor{color5}{rgb}{0.95,0.95,0}
\definecolor{color6}{rgb}{0,1,0}
\definecolor{color7}{rgb}{0.6,0.8,0.5}
\definecolor{color8}{rgb}{0.9,0.6,0}
\definecolor{color9}{rgb}{0.3,0.1,0.4}
\definecolor{color10}{rgb}{0.7,0,0}
\definecolor{color11}{rgb}{0,0.7,0}
\definecolor{color12}{rgb}{0,0,0.7}
\definecolor{color13}{rgb}{0.7,0.7,0}
\definecolor{color14}{rgb}{0.7,0,0}
\definecolor{color15}{rgb}{0,0.7,0.7}

\begin{document}

\title{Abdominal multi-organ segmentation with cascaded convolutional and adversarial deep networks}

\author{Pierre-Henri~Conze$^{\star}$, Ali Emre Kavur, Emilie Cornec-Le Gall, \textcolor{black}{Naciye} Sinem Gezer, Yannick Le Meur, M. Alper Selver and Fran\c cois Rousseau \vspace{-0.8cm}

\thanks{P.-H. Conze and F. Rousseau are with IMT Atlantique, LaTIM UMR 1101, Inserm, UBL, Brest, France.
E-mail: pierre-henri.conze@imt-atlantique.fr.
A. E. Kavur is with Graduate School of Natural and Applied Sciences, Dokuz Eylul University, Izmir, Turkey.
E. Cornec-Le Gall is with Department of Nephrology, CHRU Brest, UMR 1078, Inserm, UBO, Brest, France.
N. S. Gezer is with Department of Radiology, Faculty of Medicine, Dokuz Eylul University, Izmir, Turkey.
Y. Le Meur is with Department of Nephrology, CHRU Brest, UMR 1227, Inserm, UBO, Brest, France.
M. A. Selver is with Department of Electrical and Electronics Engineering, Dokuz Eylul University, Izmir, Turkey.
\textit{Asterisk indicates corresponding author.} Manuscript received ---- XX, 2020; revised ---- XX, 2020.}}

\markboth{}%
{Shell \MakeLowercase{\textit{et al.}}: Bare Demo of IEEEtran.cls for IEEE Journals}

\maketitle

\begin{abstract} 

Objective : Abdominal anatomy segmentation is crucial for numerous applications from computer-assisted diagnosis to image-guided surgery. In this context, we address \textcolor{black}{fully-automated} multi-organ segmentation from abdominal CT and MR images using deep learning. Methods: The proposed model extends standard conditional generative adversarial networks. Additionally to the discriminator which enforces the model to create realistic \textcolor{black}{organ} delineations, it embeds cascaded partially pre-trained convolutional encoder-decoders as generator. \textcolor{black}{Encoder fine-tuning from a large amount of non-medical images alleviates data scarcity limitations}. The network is trained end-to-end to benefit from simultaneous multi-level segmentation refinements \textcolor{black}{using auto-context}. Results : Employed for healthy liver, kidneys and spleen segmentation, our pipeline provides \textcolor{black}{promising} results \textcolor{black}{by outperforming} state-of-the-art encoder-decoder schemes. Followed for the Combined Healthy Abdominal Organ Segmentation (CHAOS) challenge organized in conjunction with the \textcolor{black}{IEEE} International Symposium on Biomedical Imaging 2019, it gave us the first rank for three competition categories: liver CT, liver MR and multi-organ MR segmentation. Conclusion : Combining cascaded convolutional and adversarial networks strengthens the ability of deep learning pipelines \textcolor{black}{to automatically delineate multiple abdominal organs}, with good generalization capability. Significance : The comprehensive evaluation provided suggests that \textcolor{black}{better guidance could} be achieved \textcolor{black}{to help clinicians in} abdominal image interpretation and clinical decision making.

\end{abstract}

\begin{IEEEkeywords}
\textcolor{black}{Multi-organ segmentation, \textcolor{black}{abdomen}, adversarial \textcolor{black}{learning}, convolutional encoder-decoders, \textcolor{black}{cascaded networks}.} 
\end{IEEEkeywords}

\IEEEpeerreviewmaketitle

\vspace{-0.4cm}

\section{Introduction}
\label{sec:sec1}

\IEEEPARstart{T}{}he development of non-invasive imaging technologies over the last decades has opened new horizons in studying abdominal anatomical structures. Segmentation has become a crucial task in abdominal image analysis with numerous applications including computer-assisted diagnosis, surgery planning (organ pre-evaluation for resection or transplantation), visual augmentation, extraction of quantitative indices or image-guided interventions \cite{summers2016progress}. In particular, the precise \textcolor{black}{delineation} of abdominal solid visceral organs including liver, kidneys and spleen for \textcolor{black}{localization, volume assessment and follow-up} purposes has critical importance. However, the analysis of Computed Tomography (CT) and Magnetic Resonance (MR) abdominal imaging datasets is \textcolor{black}{challenging} and time-consuming for \textcolor{black}{clinicians} since the abdomen is a complex body space. Robust automatic abdominal image segmentation is required to guide image interpretation, facilitate clinical decision making and improve patient care while avoiding traditional manual delineation efforts.

\begin{figure}
\begin{minipage}{1\linewidth}
\vspace{0.1cm} 
\begin{tabular}{cccc}
\hspace{-0.25cm} \includegraphics[width=2.07cm, height=1.37cm]{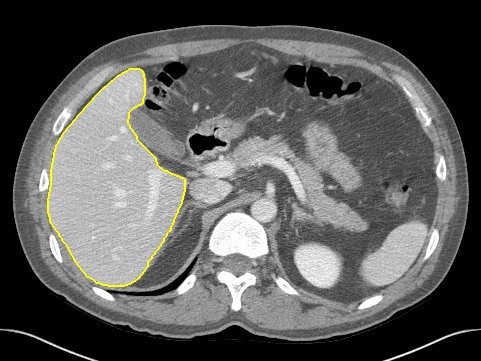} &
\hspace{-0.4cm} \includegraphics[width=2.07cm, height=1.37cm]{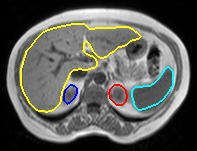} &
\hspace{-0.4cm} \includegraphics[width=2.07cm, height=1.37cm]{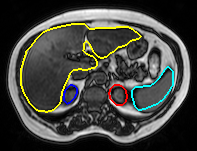} &
\hspace{-0.4cm} \includegraphics[width=2.07cm, height=1.37cm]{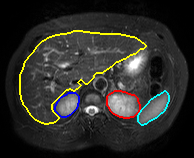} \vspace{-0.07cm} \cr
\hspace{-0.25cm} \includegraphics[width=2.07cm, height=1.37cm]{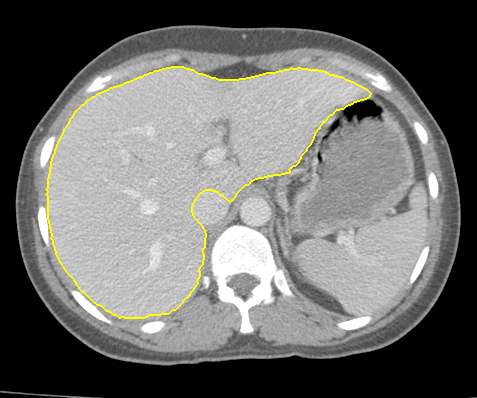} &
\hspace{-0.4cm} \includegraphics[width=2.07cm, height=1.37cm]{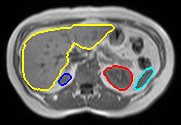} &
\hspace{-0.4cm} \includegraphics[width=2.07cm, height=1.37cm]{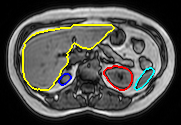} &
\hspace{-0.4cm} \includegraphics[width=2.07cm, height=1.37cm]{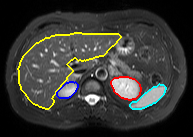} \vspace{-0.07cm} \cr
\hspace{-0.25cm} \includegraphics[width=2.07cm, height=1.37cm]{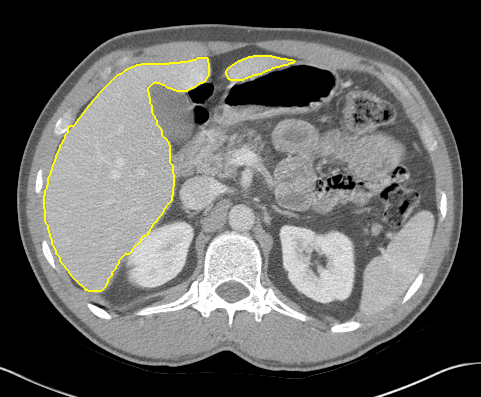} &
\hspace{-0.4cm} \includegraphics[width=2.07cm, height=1.37cm]{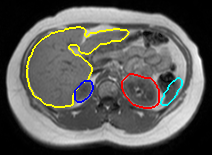} &
\hspace{-0.4cm} \includegraphics[width=2.07cm, height=1.37cm]{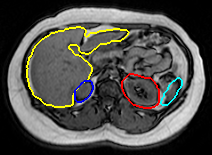} &
\hspace{-0.4cm} \includegraphics[width=2.07cm, height=1.37cm]{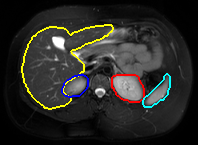} \vspace{-0.07cm} \cr
\hspace{-0.25cm} \includegraphics[width=2.07cm, height=1.37cm]{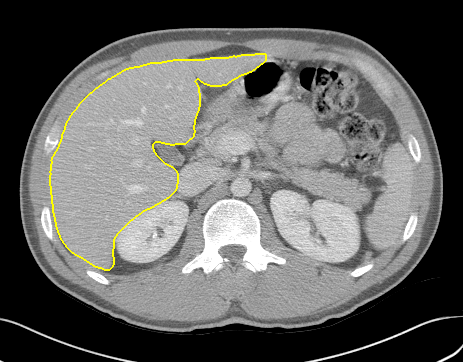} &
\hspace{-0.4cm} \includegraphics[width=2.07cm, height=1.37cm]{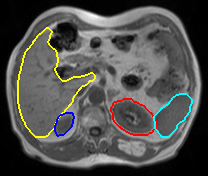} &
\hspace{-0.4cm} \includegraphics[width=2.07cm, height=1.37cm]{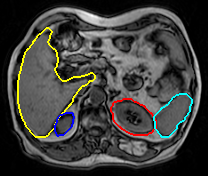} &
\hspace{-0.4cm} \includegraphics[width=2.07cm, height=1.37cm]{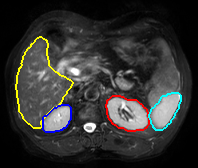} \vspace{-0.05cm} \cr
\hspace{-0.25cm} \small CT & \hspace{-0.35cm} \small T1DUAL\texttt{in} & \hspace{-0.35cm} \small T1DUAL\texttt{out} & \hspace{-0.35cm} \small T2SPIR \cr
\end{tabular} \vspace{-0.65cm} \\
\begin{center}
\textcolor{color13}{---} liver
\textcolor{color12}{---} right kidney
\textcolor{color10}{---} left kidney
\textcolor{color15}{---} spleen
\end{center}
\end{minipage}
\caption{Samples of healthy abdominal CT and MR (T1-DUAL\texttt{in/out}, T2-SPIR) images arising from the CHAOS dataset \cite{kavur2020chaos}, provided with groundtruth organ delineations.} \vspace{-0.5cm} 
\label{fig::fig-1}
\end{figure}

In this area, many interactive, semi- and fully-automated methods \textcolor{black}{have} been proposed with diverse methodologies including statistical shape models \cite{cerrolaza2015automatic}, multi-atlas segmentation \cite{xu2015efficient} or machine learning \cite{cuingnet2012automatic} techniques. More recently, outstanding performance has been reached in almost every medical image analysis tasks using deep learning \cite{litjens2017survey}. This suggests that abdominal multi-organ segmentation could further benefit from this \textcolor{black}{massive trend}, despite the large variability in abdominal organ shape, size, location and texture. Compared to conventional machine learning, the need for hand-crafted features no longer \textcolor{black}{remains} necessary. In particular, huge efforts have been devoted to automatic segmentation based on variants of Fully Convolutional Networks (FCN) \cite{long2015fully}. Recent architectures comprise a regular FCN to extract multi-scale features, followed by an up-sampling part that enables to recover the input resolution using up-convolutions \cite{litjens2017survey}. In the medical community, UNet \cite{ronneberger2015unet} is one of the most well-known approach among such Convolutional Encoder-Decoders (CED). Able to learn from relatively small datasets, CED are the most likely to automatically infer high-level knowledge involved by radiologists when interpreting abdominal images.

Despite intensive developments in deep learning, it remains difficult to judge the effectiveness of deep networks for abdominal multi-organ segmentation since they are mainly assessed on: 1- one single organ only (liver most often), 2- one single \textcolor{black}{modality} (usually CT) and 3- relatively small and private datasets. Their robustness to delineate other abdominal structures from different modalities and to manage strong inter-subject variability is therefore under- or un-investigated. Rather than organ or modality-specific strategies, the development of more comprehensive and generic computational models is needed \cite{cerrolaza2019computational}. Few challenges including the Combined Healthy Abdominal Organ Segmentation (CHAOS) challenge\footnotemark[1] \cite{kavur2020chaos}, organized in conjunction with the \textcolor{black}{IEEE} International Symposium on Biomedical Imaging (ISBI) 2019, has been proposed to motivate further work on this perspective by making available a dataset (Fig.\ref{fig::fig-1}) \textcolor{black}{to segment multiple organs from} two imaging modalities (CT, MR with T1-DUAL and T2-SPIR sequences) acquired \textcolor{black}{for} unpaired healthy subjects. Towards efficient combined segmentation and based on this unique dataset, we target robust and generic deep learning architectures for two main purposes: 1- segmentation of liver from CT scans and 2- segmentation of four abdominal organs (liver, right kidney, left kidney, spleen) from MR images.

\footnotetext[1]{\url{https://chaos.grand-challenge.org}}
\footnotetext[2]{\url{https://chaos.grand-challenge.org/results_CHAOS/}}

Our contributions dedicated to healthy abdominal multi-organ segmentation are three-fold. First, deeper CED architectures using encoders pre-trained on non-medical data and extending the UNet \cite{ronneberger2015unet} baseline are proposed. Second, we embed this architecture into a cascaded framework using auto-context and end-to-end training to benefit from simultaneous multi-level segmentation refinements. Third, such cascaded pipeline is used as generator within a conditional Generative Adversarial Network (cGAN). The resulting model thus includes a discriminator to strengthen the ability of the generative part to create delineations as realistic as possible. The step-by-step evaluation provided for each contribution in both CT and MR modalities highlights better performance than state-of-the-art encoder-decoder schemes. The pipeline also gave us the first rank for three CHAOS competition categories\footnotemark[2] (liver CT, liver MR and multi-organ MR segmentation) which suggests that the proposed computational deep models can offer new insights for abdominal image interpretation and clinical decision making in various computer-assisted tasks.

\section{Related works}
\label{sec:sec2}

Computational abdominal organ segmentation has attracted considerable attention over the last decades. This craze led to the development of a wide range of methods, from interactive to semi- and fully-automated \cite{kavur2020donors}. Before the recent development of machine and deep learning, abdominal organ segmentation has often been carried out using statistical shape models \cite{zhang2010automatic,cerrolaza2015automatic} to capture and then fit organ shapes through anatomical correspondences. Since deformations and limited datasets may prevent those models from managing the strong variability of abdominal organ shapes, aligning and merging manually segmented images could be \textcolor{black}{followed as an alternative}. \textcolor{black}{Specifically,} multi-atlas segmentation consists in leveraging label atlases through image registration and statistical fusion \cite{sabuncu2010generative}. Applied to abdominal data, coarse-to-fine \cite{yang2014automatic}, region-wise local atlas selection \cite{wolz2013automated}, Selective and Iterative Method for Performance Level Estimation (SIMPLE) \textcolor{black}{\cite{xu2015efficient,huo2017robust}} or dictionary learning and sparse coding \cite{tong2015discriminative} techniques can be employed to alleviate substantial registration errors. Nevertheless, robust inter-subject abdominal image registration is a challenging, computational intense and not yet solved issue \cite{xu2016evaluation} due to the diversity of organ shape, size, location and texture. This mainly explains the success of registration-free methods whose aim is to learn feature distributions that characterize abdominal anatomy from un-registered images.

Among registration-free methods, computational power and data availability have enabled the rise of machine learning techniques \textit{via} voxel- \cite{cuingnet2012automatic,bieth2017large}, \textcolor{black}{patch- \cite{giraud2016optimized}} or supervoxel-wise \cite{conze2017ijcars} classifiers. These methods require hand-crafted features and therefore, specialized knowledge to delineate anatomical structures. Conversely, deep Convolutional Neural Networks (CNN), data-driven supervised learning models formed by multi-layer neural networks, automatically learn complex hierarchical features from data \cite{lecun1998gradient}. In this direction, huge efforts have been devoted to automatic segmentation based on variants of Fully Convolutional Networks (FCN) \cite{long2015fully}. Further improvements are reached with architectures comprising a regular FCN to extract features, followed by an up-sampling part which recovers the input resolution through up-convolutions \cite{litjens2017survey}. In the medical community, UNet \cite{ronneberger2015unet} and its 3D counterparts \cite{milletari2016vnet,cciccek20163d} are among the most well-known Convolutional Encoder-Decoders (CED). They exploit \textit{skip connections} \textcolor{black}{to} concatenate features between contracting and expanding paths \textcolor{black}{for improving} localization accuracy while allowing faster convergence.

CED networks have been widely adopted for automatic abdominal organ segmentation, as in \cite{roth2017hierarchical} where 3D UNet \cite{cciccek20163d} is exploited in a two-stage hierarchical fashion for multi-organ delineation purposes. Combining densely linked layers and shallow 3D UNet architecture \cite{gibson2018automatic} enables high-resolution activation maps through memory-efficient dropout and feature re-use. Some approaches consider post-processing steps for further contour refinement by  exploiting organ probability maps arising from 3D CED as features for Conditional Random Field (CRF) \cite{dou20163d}, level-set \cite{hu2017automatic} or graph-cut \cite{lu2017automatic} models. Organ-attention networks with reverse connections followed by statistical fusion \cite{wang2019abdominal} tend to \textcolor{black}{reduce} uncertainties at weak boundaries and deal with relative organ size variations. 

Feeding deep networks with volumetric images obviously faces memory and computational issues. Since increasing the network depth to extract discriminative features with a larger receptive field cannot be done \textit{ad-infinitum}, many methods rely on small patches or downsampled images resulting in a significant loss of spatial context \cite{gibson2018automatic}. Reaching accurate abdominal organ delineations\textcolor{black}{, however,} requires to extract high-level contextual information, as do radiologists visually. Several key contributions in semantic segmentation \textcolor{black}{arose to mimic} visual medical image interpretation \textcolor{black}{more closely}. First, structure delineation can exploit transfer learning from large non-medical datasets \cite{iglovikov2018ternausnet,conze19isbi} to reduce the data scarcity issue while improving model generalizability \cite{yosinski2014transferable}. Second, stacking multiple CEDs encourages the integration of more representative multi-level information \cite{choi2016fast, roth18pyramid}. In particular, cascades of deep CEDs can embed auto-context \cite{tu2010auto} to fuse various amounts of spatial context by using posterior probabilities resulting from one CED block to the subsequent. Third, conditional \textcolor{black}{g}enerative \textcolor{black}{a}dversarial \textcolor{black}{n}etworks extends standard image-to-image translation \textcolor{black}{\cite{isola2017image}} by including a discriminator whose role is to enforce the model to generate realistic outputs. These avenues represent promising methodological developments to achieve more generic computational models \textcolor{black}{for} CT and MRI abdominal multi-organ \textcolor{black}{segmentation}. 

\section{Methods}
\label{sec:sec3}

\subsection{Conditional generative adversarial networks}
\label{sec:ssec3-1}

\begin{figure}
\vspace{-0.2cm}
\hspace{-1.5cm} \begin{center}
\begin{tikzpicture}[thick]
\tikzstyle{every node} = [minimum width=20pt, minimum height=10pt]
\node(l1)[] at (0,1) {\scriptsize source image };
\node(u1)[draw, rectangle] at (0,0) {\includegraphics[height=1.4cm]{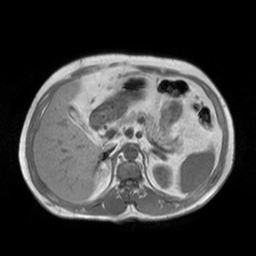} };
\node(u2)[draw, circle] at (2.5,0) {\begin{minipage}{0.18\linewidth}\hspace{0.04cm} \vspace{0.3cm} generator\end{minipage}};
\node(l1)[] at (4.9,1) {\scriptsize predicted };
\node(u3)[draw, rectangle] at (4.9,0) {\includegraphics[height=1.4cm]{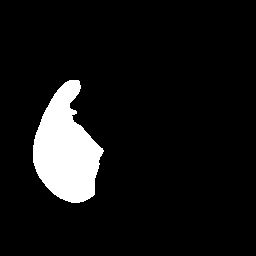} };
\node(l1)[] at (7,1) {\scriptsize groundtruth };
\node(u4)[draw, rectangle] at (7,0) {\includegraphics[height=1.4cm]{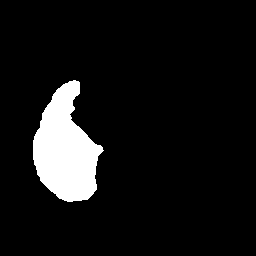} };
\node(u5)[draw, circle] at (4.9,-2.5) {\begin{minipage}{0.22\linewidth} \hspace{-0.01cm} \vspace{0.4cm} discriminator\end{minipage}};
\node(u6)[draw, rectangle] at (7,-4.3) {\small \texttt{dice} loss};
\node(u7)[draw, rectangle] at (4.9,-4.3) {\small \texttt{BCE} loss};
\node(u8)[draw] at (2.5,-4.3) {\textcircled{$+$}};
\node(a1)[] at (0.22,-1.1) {\textcolor{black}{$x$}};
\node(a2)[] at (4.45,-1.1) {\textcolor{black}{$G(x)$}};
\node(a3)[] at (6.7,-1.1) {\textcolor{black}{$y$}};
\node(a4)[] at (2.5,-0.26) {\textcolor{black}{$G$}};
\node(a5)[] at (4.9,-2.75) {\textcolor{black}{$D$}};
\node(a6)[] at (1.73,-3.6) {\textcolor{black}{$l_{G}(G,D)$}};
\path[->,draw,>=latex]
(u1) edge (u2)
(u1) edge[in=-175, out=-100] (u5)
(u2) edge (u3)
(u8) edge (u2)
(u6) edge[in=-20, out=-100] (u8)
(u7) edge (u8)
(u3) edge (u5)
(u4) edge (u5)
(u3) edge (u6)
(u4) edge (u6)
(u5) edge (u7);
\end{tikzpicture}  \vspace{-1cm}
\end{center}
\caption{Conditional generative adversarial networks combining \texttt{dice} and \texttt{BCE} losses for abdominal organ segmentation.} \vspace{-0.3cm} 
\label{fig::fig-2}
\end{figure}
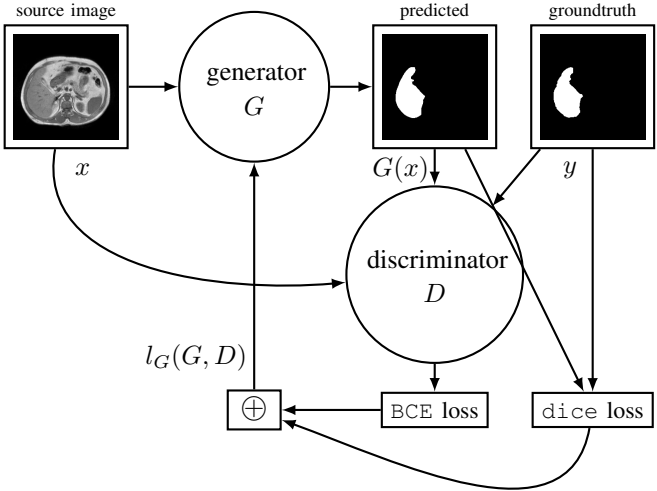

\begin{figure}
\hspace{0.1cm} \begin{minipage}{0.5\linewidth}
\includegraphics[width=8cm]{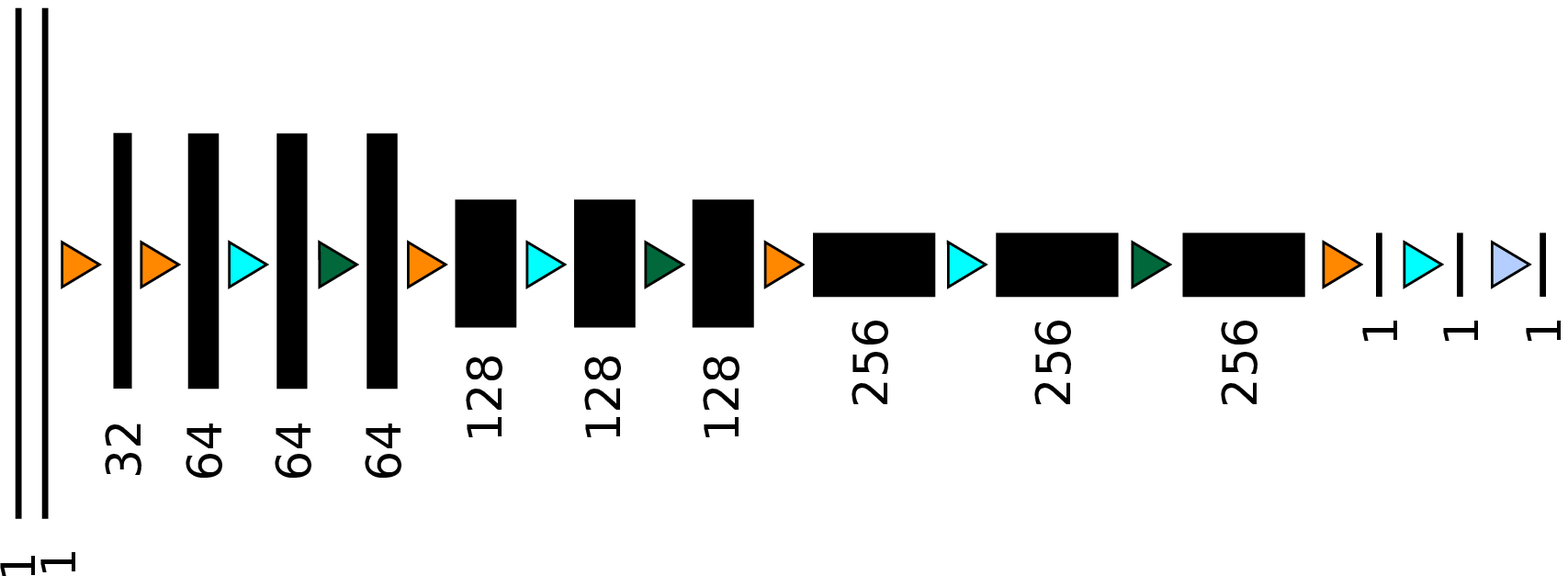} \vspace{-0.15cm} \\
\includegraphics[width=8cm]{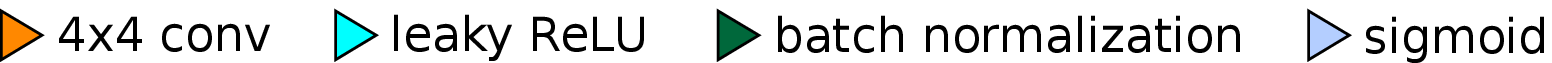}
\end{minipage}
\caption{Discriminative part of conditional generative adversarial networks.} \vspace{-0.5cm} 
\label{fig::fig-3}
\end{figure}

Recent works including \textcolor{black}{\cite{singh2020breast,boutillon2020combining}} have demonstrated the feasibility \textcolor{black}{of} image-to-image \textcolor{black}{translation} \cite{isola2017image} based on conditional Generative Adversarial Networks (cGAN) for medical image segmentation purposes. cGAN architectures (Fig.\ref{fig::fig-2}) are made of a generator which provides segmentation masks through encoding and decoding layers and a discriminator (Fig.\ref{fig::fig-3}) which assesses if a given segmentation mask is synthetic or real. Thus, the adversarial network learns to discriminate real (groundtruth) from synthetic delineations (those arising from the generator) to enforce the generative part to create segmentation masks as plausible as possible.

\begin{figure*}
\vspace{-0.1cm}
\begin{tabular}{lc}
\rotatebox{90}{\small \hspace{0.45cm} (a) \texttt{UNet}} & \includegraphics[width=16.4cm]{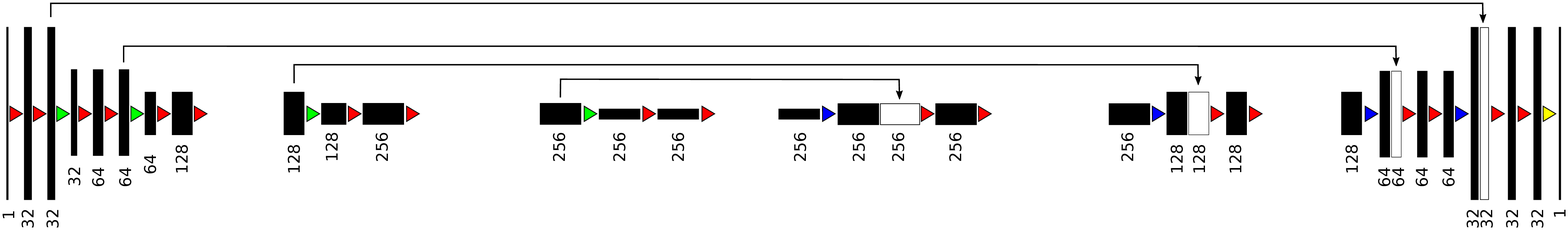} \cr
\rotatebox{90}{\small \hspace{0.17cm} (b) \texttt{v19UNet}} & \includegraphics[width=16.4cm]{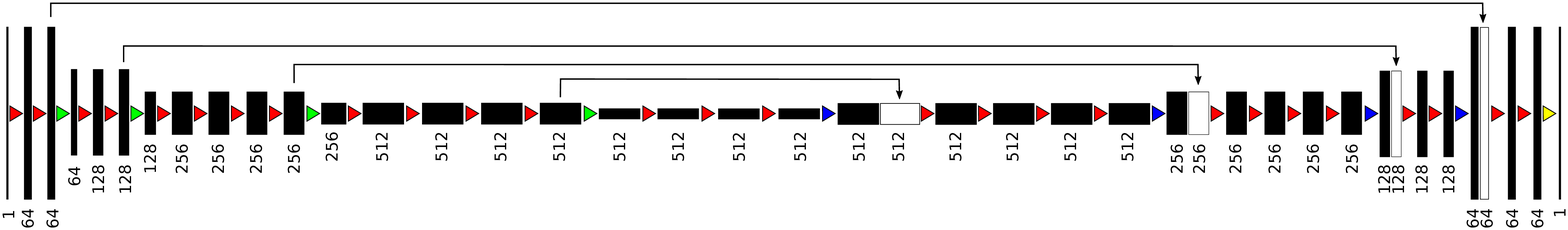} \cr
\rotatebox{90}{\small \hspace{0.1cm} (c) \texttt{v19pUNet}} & \includegraphics[width=16.4cm]{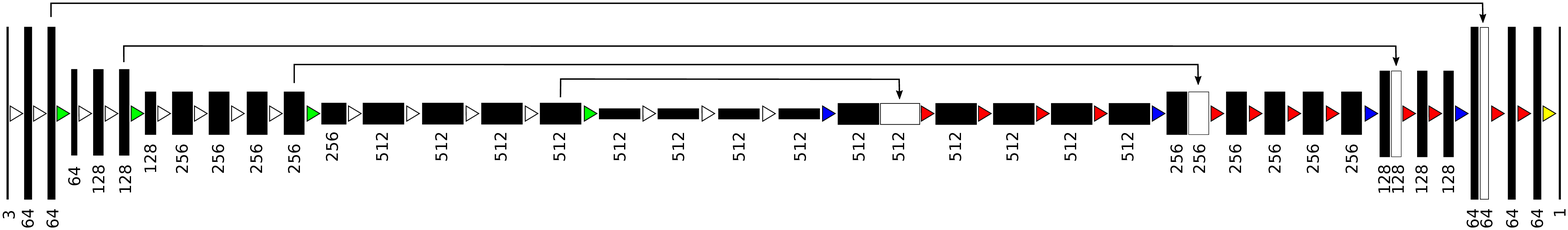} \cr
& \includegraphics[width=16.4cm]{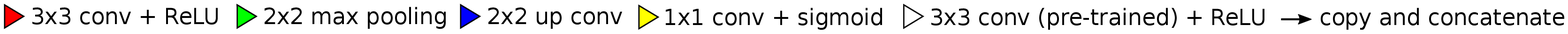} \cr
\end{tabular} \vspace{-0.05cm}
\caption{Extension of UNet \cite{ronneberger2015unet} (a) by exploiting as encoder a slightly modified VGG-19 \cite{simonyan2014very} without (b) and with (c) weights pre-trained on ImageNet \cite{russakovsky2015imagenet}. The decoder is modified to get \textcolor{black}{a symmetrical} construction while keeping \textit{skip connections}.} \vspace{-0.475cm} 
\label{fig::fig-4}
\end{figure*} 

cGAN pipelines usually use UNet as generator $G$ (Fig.\ref{fig::fig-4}\textit{a}). Its symmetrical architecture comprises an encoder which gradually reduce the spatial dimension using pooling layers, a decoder progressively recovering object details and initial resolution as well as \textit{skip connections} which concatenate features between contracting and expanding paths. \textcolor{black}{Specifically,} UNet consists of sequential layers including $3\hspace{-0.03cm}\times\hspace{-0.03cm}3$ convolutional layers followed by Rectified Linear Unit (ReLU) activations. Spatial size is reduced using $2\hspace{-0.01cm}\times\hspace{-0.01cm}2$ max pooling layers. The first convolutional layer generates 32 channels \textcolor{black}{\cite{ronneberger2015unet}}. This number doubles after each pooling as the network deepens. \textcolor{black}{Following \cite{singh2020breast},} the discriminator $D$ consists of five $4\hspace{-0.09cm}\times\hspace{-0.09cm}4$ convolutional layers followed by leaky ReLU activation functions and batch normalization (Fig.\ref{fig::fig-3}). The discriminator inputs are the concatenation of both \textcolor{black}{source} images and groundtruth or predicted binary masks to be evaluated. The output is an array where each value is defined between $0$ (fake) and $1$ (plausible or real) and corresponds to the degree of segmentation likelihood for a \textcolor{black}{given} image and segmentation mask crop. Let $x$ and $y$ be the source images and groundtruth delineation masks, $\lambda \hspace{0.1cm}\textcolor{black}{=150}$ a weighting factor, $G(x)$ and $D(x,G(x))$ the outputs of $G$ and $D$, $l_{\texttt{dice}}$ the Dice loss estimated by comparing predicted and groundtruth masks. The loss function for $G$ is \textcolor{black}{as follows}: \vspace{-0.225cm} 

\begin{equation}
l_{G}(G,D) = \underset{x,y}{\mathbb{E}}[-\texttt{log}(D(x,G(x)))] + \lambda \hspace{0.05cm} \underset{x,y}{\mathbb{E}}[l_{\texttt{dice}}(G(x),y)]
\label{eq::eq-1} \vspace{0.15cm} 
\end{equation}

\noindent Minimizing $l_{\texttt{dice}}$ tends to provide rough \textcolor{black}{organ shape} predictions whereas maximizing $\texttt{log}(D(x,G(x))$ aims at improving contour delineations. The loss function for $D$ is \textcolor{black}{such} that:  \vspace{-0.25cm} 

\begin{eqnarray}
l_{D}(G,D) &=& \underset{x,y}{\mathbb{E}}[-\texttt{log}(D(x,y))] \nonumber \\
&+& \underset{x,y}{\mathbb{E}}[-\texttt{log}(1-D(x,G(x)))]
\label{eq::eq-2}
\end{eqnarray} \vspace{-0.2cm}

\noindent The optimizer fits $D$ through Binary Cross Entropy (\texttt{BCE}) using estimated and grountruth masks. It maximizes loss values for groundtruth ($\texttt{log}(D(x,y))$) and minimizes loss values for generated ($-\texttt{log}(1-D(x,G(x)))$) masks. Optimization is performed sequentially \textcolor{black}{by alternating at each batch gradient descents on $G$ and $D$ \cite{goodfellow2014generative}}. \textcolor{black}{To further improve cGAN abilities to extract contours from the abdominal anatomy, investigations on more robust generators than traditional UNet are needed.} \vspace{-0.05cm}

\subsection{Partially pre-trained generator}
\label{sec:ssec3-2}

CED architectures dedicated to medical image segmentation are typically trained from scratch \textit{via} randomly initialized weights. \textcolor{black}{Since the amount of available images cannot be endlessly extended,} reaching a generic model without over-fitting is therefore \textcolor{black}{challenging}. \textcolor{black}{As deep classification networks which usually involve model pre-trained on large datasets}, the encoder part of \textcolor{black}{CEDs} can be replaced by a classification network whose weights are \textcolor{black}{previously trained} on an initial classification task \textcolor{black}{\cite{iglovikov2018ternausnet}}. It \textcolor{black}{exploits} transfer learning and fine-tuning from large datasets \textcolor{black}{like} ImageNet \cite{russakovsky2015imagenet} towards \textcolor{black}{better semantic} segmentation. In the literature, the encoder has been already replaced \textcolor{black}{by} pre-trained VGG-11 \cite{iglovikov2018ternausnet} or WideResnet-38 \cite{iglovikov2018ternausnetv2} \textcolor{black}{networks}. Our previous study \cite{conze19isbi} \textcolor{black}{exploits} pre-trained VGG-16 encoders and \textcolor{black}{reveals} significant improvements compared to their randomly weighted counterparts. 

This approach can be further improved by extending standard UNet \cite{ronneberger2015unet} by a deeper network from the VGG family \cite{simonyan2014very} as encoder: the VGG-19 architecture. Compared to UNet (Fig.\ref{fig::fig-4}\textit{a}), the first convolutional layer of \texttt{v19UNet} (Fig.\ref{fig::fig-4}\textit{b}) generates $64$ channels instead of $32$. The number of channels doubles after each max pooling until it reaches $512$ ($256$ for UNet). After the second max pooling, the number of convolutional layers differs from UNet with patterns of $4$ consecutive layers instead of $2$. Compared to VGG-19 \textcolor{black}{\cite{simonyan2014very}}, top layers including fully-connected layers and softmax are omitted. The three last convolutional VGG-19 layers serve as central part to separate contracting and expanding paths. To improve performance, this encoder branch is pre-trained on ImageNet \cite{russakovsky2015imagenet} to get \texttt{v19pUNet} (Fig.\ref{fig::fig-4}\textit{c}). Pre-training this encoder using more than $1$ million non-medical data collected for object recognition purposes \textcolor{black}{improves} predictive performance on abdominal data \textcolor{black}{and requires} less training time to reach convergence. In practice, axial slices are extended from single greyscale channel to $3$ channels by repeating the same content \textcolor{black}{to fit the RGB ImageNet image dimensions}.

\begin{figure}
\vspace{-0.25cm}
\hspace{-1.5cm} \begin{center}
\begin{tikzpicture}[thick]
\tikzstyle{every node} = [minimum width=20pt, minimum height=10pt]
\node(u0a)[draw, rectangle] at (0,0) {\includegraphics[height=1.4cm]{./visu-scheme/02-z013-src-in.png} };
\node(l1)[] at (0,1.05) {\scriptsize source image };
\node(u2c)[draw, rectangle] at (2.2,0) {\small stack};
\node(u3)[draw, circle] at (4.4,0) {\begin{minipage}{0.17\linewidth} \small \texttt{v19pUNet} \\ \tiny \textcolor{white}{--} \textcolor{black}{linear activation} \end{minipage}};
\node(u4a)[draw, rectangle] at (6.9,0) {\includegraphics[height=1.4cm]{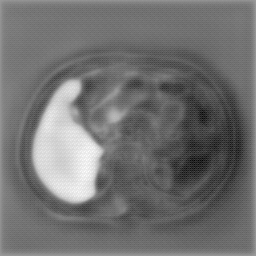}};
\node(l2)[] at (6.9,1.05) {\scriptsize posterior prob. };
\node(u6a)[draw, rectangle] at (0,-2.9) {\includegraphics[height=1.4cm]{./visu-scheme/02-z013-pred.png}};
\node(l3)[] at (0.03,-1.85) {\scriptsize posterior prob. };
\node(u7)[draw, rectangle] at (1.7,-2.9) {\small stack};
\node(u8)[draw, circle] at (3.6,-2.9) {\begin{minipage}{0.17\linewidth} \small \texttt{v19pUNet} \\ \tiny \textcolor{white}{-} \textcolor{black}{sigmoid activation} \end{minipage}};
\node(u10)[draw, rectangle] at (5.3,-2.9) {\rotatebox{90}{\small post-processing}};
\node(u11a)[draw, rectangle] at (6.9,-2.9) {\includegraphics[height=1.4cm]{./visu-scheme/02-z013-pred-3.png}};
\node(l4)[] at (6.9,-1.85) {\scriptsize predicted };
\path[->=stealth',->]
(u0a) edge[bend right=20] (u2c)
(u0a) edge[bend left=20] (u2c)
(u0a) edge (u2c)
(u2c) edge (u3)
(u3) edge (u4a)
(u4a) edge[in=100, out=-130] (u6a)
(u6a) edge (u7)
(u0a) edge[in=90, out=-35] (u7)
(u0a) edge[in=-90, out=-74] (u7)
(u7) edge (u8)
(u8) edge (u10)
(u10) edge (u11a);
\end{tikzpicture} \vspace{-0.5cm}
\end{center}
\caption{Cascaded convolutional encoder-decoders with auto-context to exploit multi-level contextual information.} \vspace{-0.525cm} 
\label{fig::fig-5}
\end{figure}

To get a \textcolor{black}{symmetrical} construction while keeping \textit{skip connections} (Fig.\ref{fig::fig-4}\textit{c}), \textcolor{black}{the decoder branch is extended in the same fashion} by adding $4$ convolutional layers \textcolor{black}{and} more features channels. Contrary to encoder weights which are initialized \textcolor{black}{\textit{via} ImageNet pre-training}, decoder weights are set randomly. \textcolor{black}{A} final $1\hspace{-0.05cm}\times\hspace{-0.05cm}1$ convolutional layer \textcolor{black}{with} sigmoid activation function achieves pixel-wise \textcolor{black}{segmentation} at native resolution. \vspace{-0.3cm}

\subsection{Cascaded generator with auto-context}
\label{sec:ssec3-3}

Managing long-range spatial context is key to improve \textcolor{black}{abdominal} organ delineations \textcolor{black}{\cite{conze2017ijcars}}. However, increasing \textit{ad-infinitum} the network depth to exploit larger receptive fields is not suitable \textcolor{black}{for} memory and computational issues. Alternatively, in the same spirit of \cite{roth18pyramid}, we propose to process abdominal images using a cascade of deep CEDs to exploit multi-level contextual information (Fig.\ref{fig::fig-5}). Our strategy, referred to \texttt{v19pUNet\tiny{1-1}}, consists in combining two \textcolor{black}{partially pre-trained} \texttt{v19pUNet} networks with auto-context \cite{tu2010auto}, i.e. using posterior probabilities resulting from the first \texttt{v19pUNet} as features for the second one \cite{salehi2017auto}. \textcolor{black}{It extends with more complex architectures a proof-of concept we gave in \cite{yan19cascaded} using standard UNet (\texttt{UNet\tiny{1-1}})}. The sigmoid activation of the first \texttt{v19pUNet} used in the last $1\hspace{-0.08cm}\times\hspace{-0.08cm}1$ convolution layer (Fig.\ref{fig::fig-4}\textit{c}) is replaced by a linear function to generate continuous output maps. These maps are normalized, concatenated to source images and given as inputs of the second \texttt{v19pUNet} which is trained to \textcolor{black}{give} final organ delineations. Instead of training both models separately \cite{salehi2017auto}, our pipeline is trained end-to-end to exploit simultaneous multi-level segmentation refinements. Making the first \texttt{v19pUNet} generating continuous instead of binary outputs \textcolor{black}{propagates} pixel-wise confidence information to the second \texttt{v19pUNet} and postpones the final segmentation decision to the pipeline ending part. Contrary to \cite{roth18pyramid}, both networks process source images at full-resolution. Moreover, we keep the largest connected segmented area as post-processing.

\usetikzlibrary{spy,calc}

\begin{figure*}
\scriptsize
\vspace{-0.2cm}
\hspace{-0.2cm} \begin{minipage}{0.5\linewidth}
\begin{tabular}{ccccc}
\hspace{-0.2cm} \begin{tikzpicture}[spy using outlines={rectangle,white,magnification=2,size=0.9cm, connect spies}]
\node {\includegraphics[height=2.61cm]{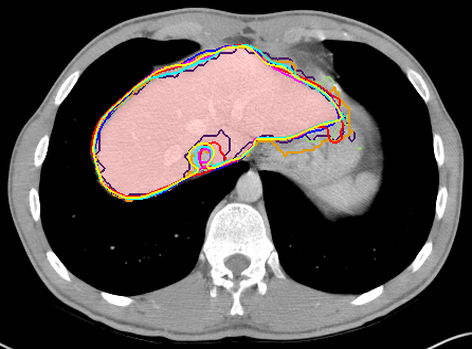}};
\spy on (-0.22,0.17) in node [left] at (1.73,-0.82);
\end{tikzpicture} &
\hspace{-0.57cm} \begin{tikzpicture}[spy using outlines={rectangle,white,magnification=2,size=0.9cm, connect spies}]
\node {\includegraphics[height=2.61cm]{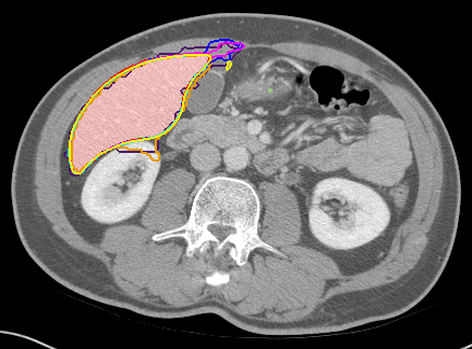}};
\spy on (-0.15,0.9) in node [left] at (1.73,-0.82);
\end{tikzpicture} &
\hspace{-0.57cm} \begin{tikzpicture}[spy using outlines={rectangle,white,magnification=2,size=0.9cm, connect spies}]
\node {\includegraphics[height=2.61cm]{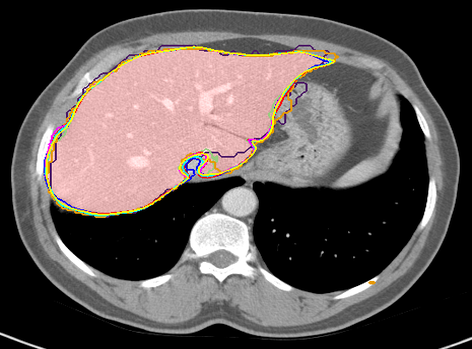}};
\spy on (-0.3,0.05) in node [left] at (1.73,-0.82);
\end{tikzpicture} &
\hspace{-0.57cm} \begin{tikzpicture}[spy using outlines={rectangle,white,magnification=2,size=0.9cm, connect spies}]
\node {\includegraphics[height=2.61cm]{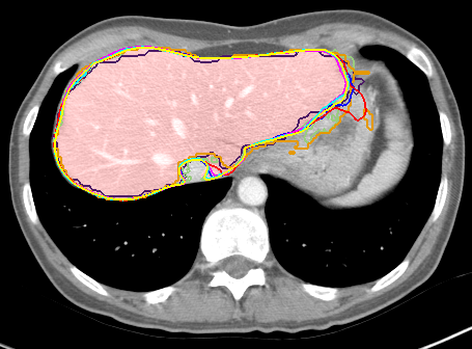}};
\spy on (-0.28,0.03) in node [left] at (-0.83,-0.82);
\spy on (0.77,0.6) in node [left] at (1.73,-0.82);
\end{tikzpicture} &
\hspace{-0.57cm} \begin{tikzpicture}[spy using outlines={rectangle,white,magnification=2,size=1.3cm, connect spies}]
\node {\includegraphics[height=2.61cm]{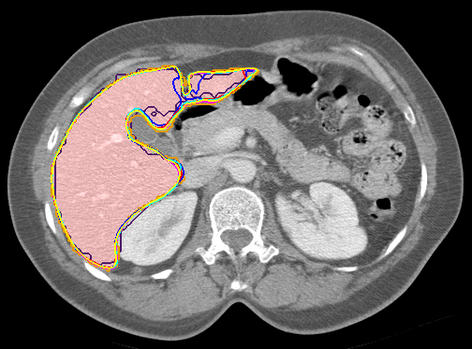}};
\spy on (-0.14,0.68) in node [left] at (1.73,-0.62);
\end{tikzpicture} \vspace{-0.15cm} \cr
\hspace{-0.2cm} \begin{tikzpicture}[spy using outlines={rectangle,white,magnification=2,size=0.9cm, connect spies}]
\node {\includegraphics[height=2.61cm]{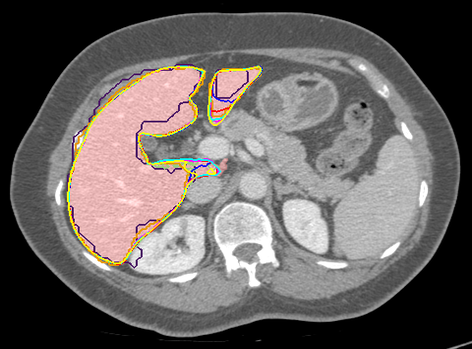}};
\spy on (-0.25,0.0) in node [left] at (1.73,-0.82);
\end{tikzpicture} &
\hspace{-0.57cm} \begin{tikzpicture}[spy using outlines={rectangle,white,magnification=2,size=1.1cm, connect spies}]
\node {\includegraphics[height=2.61cm]{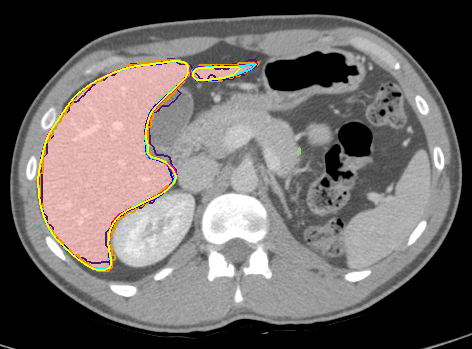}};
\spy on (-0.08,0.78) in node [left] at (1.73,-0.72);
\end{tikzpicture} &
\hspace{-0.57cm} \begin{tikzpicture}[spy using outlines={rectangle,white,magnification=2,size=0.9cm, connect spies}]
\node {\includegraphics[height=2.61cm]{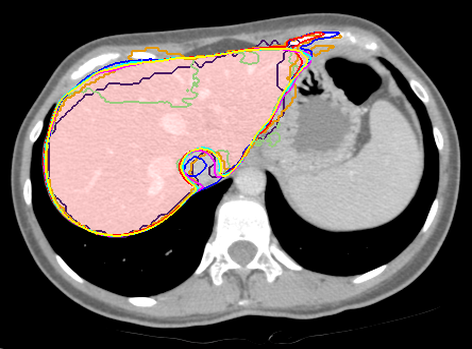}};
\spy on (-0.24,0.04) in node [left] at (1.73,-0.82);
\spy on (0.55,0.92) in node [left] at (1.73,0.13);
\end{tikzpicture} &
\hspace{-0.57cm} \begin{tikzpicture}[spy using outlines={rectangle,white,magnification=2,size=0.9cm, connect spies}]
\node {\includegraphics[height=2.61cm]{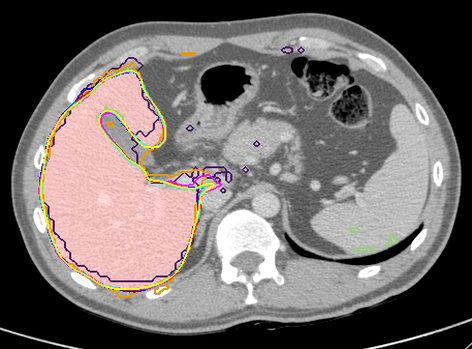}};
\spy on (-0.25,-0.11) in node [left] at (1.73,-0.82);
\end{tikzpicture} &
\hspace{-0.66cm} \begin{tikzpicture}[spy using outlines={rectangle,white,magnification=2,size=1cm, connect spies}]
\node {\includegraphics[height=2.61cm]{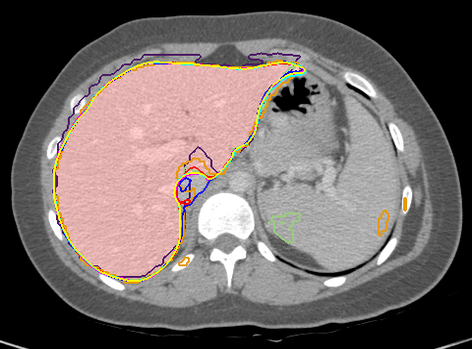}};
\spy on (-0.32,-0.025) in node [left] at (1.73,-0.77);
\end{tikzpicture} \cr 
\end{tabular}
\end{minipage} \vspace{0.05cm} \\
\textcolor{color7}{------} \texttt{DeepMedic} \cite{kamnitsas2017efficient}
\textcolor{color9}{------} \texttt{VNet} \cite{milletari2016vnet}
\textcolor{color8}{------} \texttt{denseVNet} \cite{gibson2018automatic}
\textcolor{color1}{------} \texttt{UNet} \cite{ronneberger2015unet}
\textcolor{color2}{------} \texttt{v16UNet}
\textcolor{color3}{------} \texttt{v16pUNet} \cite{conze19isbi}
\textcolor{color4}{------} \texttt{v16pUNet\tiny{1-1}}
\textcolor{color5}{------} \texttt{cGv16pUNet\tiny{1-1}}
\textcolor{color14}{$\blacksquare$} liver
\caption{Liver CT segmentation using \texttt{DeepMedic} \cite{kamnitsas2017efficient}, \texttt{VNet} \cite{milletari2016vnet}, \texttt{denseVNet} \cite{gibson2018automatic}, \texttt{UNet} \cite{ronneberger2015unet}, \texttt{v16UNet}, \texttt{v16pUNet} \cite{conze19isbi}, proposed \texttt{v16pUNet\tiny{1-1}} and \texttt{cGv16pUNet\tiny{1-1}}. Groundtruth is superimposed in red color.} \vspace{-0.2cm}
\label{fig::fig-6}
\end{figure*}

\begin{figure*}
\scriptsize
\hspace{-0.2cm} \begin{minipage}{0.5\linewidth}
\begin{tabular}{cccccc}
\rotatebox{90}{\textcolor{white}{------} T1-DUAL\texttt{in}} &
\hspace{-0.3cm} \includegraphics[height=2.63cm]{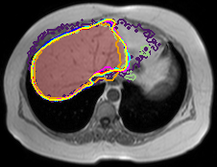} &
\hspace{-0.4cm} \includegraphics[height=2.63cm]{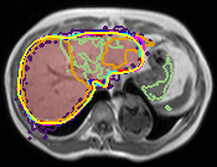} &
\hspace{-0.4cm} \includegraphics[height=2.63cm]{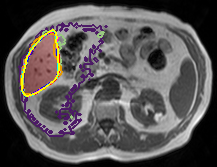} &
\hspace{-0.4cm} \includegraphics[height=2.63cm]{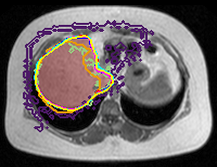} &
\hspace{-0.4cm} \includegraphics[height=2.63cm]{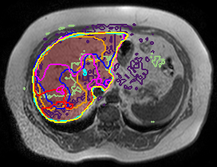} \cr
\rotatebox{90}{\textcolor{white}{----------} T2-SPIR} &
\hspace{-0.3cm} \includegraphics[height=2.63cm]{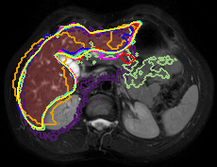} &
\hspace{-0.4cm} \includegraphics[height=2.63cm]{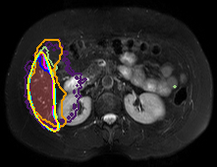} &
\hspace{-0.4cm} \includegraphics[height=2.63cm]{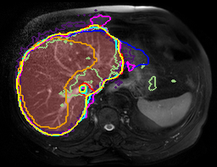} &
\hspace{-0.4cm} \includegraphics[height=2.63cm]{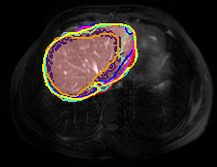} &
\hspace{-0.4cm} \includegraphics[height=2.63cm]{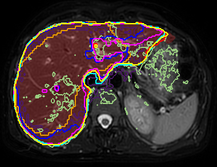} \cr
\end{tabular}
\end{minipage} \vspace{0.15cm} \\
\textcolor{color7}{------} \texttt{DeepMedic} \cite{kamnitsas2017efficient}
\textcolor{color9}{------} \texttt{VNet} \cite{milletari2016vnet}
\textcolor{color8}{------} \texttt{denseVNet} \cite{gibson2018automatic}
\textcolor{color1}{------} \texttt{UNet} \cite{ronneberger2015unet}
\textcolor{color2}{------} \texttt{v19pUNet}
\textcolor{color3}{------} \texttt{v19pUNet$+$} \cite{zhou2018unet}
\textcolor{color4}{------} \texttt{v19pUNet\tiny{1-1}}
\textcolor{color5}{------} \texttt{cGv19pUNet\tiny{1-1}}
\textcolor{color10}{$\blacksquare$} liver
\caption{Liver MRI (T1-DUAL\texttt{in}, T2-SPIR) segmentation using  \texttt{DeepMedic} \cite{kamnitsas2017efficient}, \texttt{VNet} \cite{milletari2016vnet}, \texttt{denseVNet} \cite{gibson2018automatic}, \texttt{UNet} \cite{ronneberger2015unet}, \texttt{v19pUNet}, \texttt{v19pUNet$+$} \cite{zhou2018unet}, proposed \texttt{v19pUNet\tiny{1-1}} and \texttt{cGv19pUNet\tiny{1-1}}. Groundtruth is superimposed in red color.} \vspace{-0.3cm}
\label{fig::fig-7}
\end{figure*}

We propose to use this cascaded partially pre-trained \texttt{v19pUNet\tiny{1-1}} model as generator within the cGAN pipeline (\texttt{cGv19pUNet\tiny{1-1}}). \textcolor{black}{Robustness and generalization capabilities} need to be assessed for abdominal multi-organ segmentation.

\section{Results and discussion}
\label{sec:sec4}

\subsection{Validation setup}
\label{sec:ssec4-1}

Results are provided using the dataset\textcolor{black}{\footnotemark[3]} arising from the CHAOS challenge \cite{kavur2020chaos}, collected by the Department of Radiology, Dokuz Eylul University Hospital, Izmir, Turkey. 80 patients are involved. 40 abdominal CT scans acquired at portal venous phase after contrast agent injection are used with groundtruth liver segmentation. The dataset also includes 40 T1-DUAL in phase (T1-DUAL\texttt{in}), 40 T1-DUAL oppose phase (T1-DUAL\texttt{out}) and 40 T2-SPIR abdominal MR images with groundtruth delineations for liver, right kidney, left kidney and spleen. Three radiology experts (10, 12 and 28 years of experience) were involved for manual segmentation. Final groundtruth masks were obtained through majority voting. T1-DUAL\texttt{in} and T1-DUAL\texttt{out} images are registered. Conversely, T1-DUAL and T2-SPIR sequences are not registered. Following the CHAOS challenge rules, CT and MR datasets are divided into training and test subsets, with a ratio of 50\%. 

\begin{table*}
\scriptsize 
\centering \begin{tabular}{|c|l|ccccc|ccccc|}
\hline
\multirow{2}{*}{organ} & \multirow{2}{*}{\hspace{-0.1cm}model} & \multicolumn{5}{c|}{CT} & \multicolumn{5}{c|}{MRI} \\ \cline{3-12}
& & \texttt{dice} \textcolor{black}{$\uparrow$} & \texttt{RAVD} \textcolor{black}{$\downarrow$} & \texttt{ASSD} \textcolor{black}{$\downarrow$} & \texttt{MSSD} \textcolor{black}{$\downarrow$} & \textbf{score} & \texttt{dice} \textcolor{black}{$\uparrow$} & \texttt{RAVD} \textcolor{black}{$\downarrow$} & \texttt{ASSD} \textcolor{black}{$\downarrow$} & \texttt{MSSD} \textcolor{black}{$\downarrow$} & \textbf{score} \\ \hline
\multirow{15}{*}{\rotatebox{90}{liver}} & \hspace{-0.2cm} \texttt{DeepMedic} \cite{kamnitsas2017efficient} \hspace{-0.2cm} & 96.70$\pm$1.36 & 3.18$\pm$3.42 & 1.24$\pm$0.48 & 27.90$\pm$10.0 & 73.32 & 89.74$\pm$7.54 & 6.52$\pm$8.27 & 4.74$\pm$4.83 & 122.5$\pm$53.2 & 47.64 \\
& \hspace{-0.2cm} \texttt{VNet} \cite{milletari2016vnet} \hspace{-0.2cm} & 89.58$\pm$8.54 & 6.78$\pm$12.6 & 4.87$\pm$8.80 & 42.52$\pm$48.6 & 60.01 & 74.55$\pm$6.23 & 42.5$\pm$26.2 & 9.21$\pm$3.63 & 75.59$\pm$31.2 & 16.81 \\
& \hspace{-0.2cm} \texttt{denseVNet} \cite{gibson2018automatic} \hspace{-0.2cm} & 95.26$\pm$1.14 & 2.89$\pm$2.53 & 1.57$\pm$0.48 & 23.89$\pm$9.19 & 73.78 & 76.75$\pm$6.86 & 17.4$\pm$13.1 & 8.27$\pm$3.12 & 54.98$\pm$28.4 & 28.91 \\
& \hspace{-0.2cm} \texttt{UNet} \cite{ronneberger2015unet} \hspace{-0.2cm} & 97.35$\pm$0.50 & 1.80$\pm$1.35 & 1.09$\pm$0.46 & 22.72$\pm$10.6 & 79.07 & 90.68$\pm$5.30 & 7.89$\pm$8.69 & 3.29$\pm$2.39 & 44.49$\pm$24.0 & 58.02 \\
& \hspace{-0.2cm} \texttt{v16UNet} \hspace{-0.2cm} & 97.67$\pm$0.41 & 1.39$\pm$1.15 & 0.88$\pm$0.25 & 19.85$\pm$8.92 & 82.71 & 91.60$\pm$5.44 & 6.87$\pm$8.63 & 2.96$\pm$2.37 & 40.75$\pm$25.2 & 60.86 \\
& \hspace{-0.2cm} \texttt{v16pUNet} \cite{conze19isbi} \hspace{-0.2cm} & 97.86$\pm$0.32 & 1.29$\pm$1.01 & 0.80$\pm$0.24 & 19.09$\pm$8.84 & 83.71 & 94.07$\pm$2.32 & 4.25$\pm$3.46 & \underline{1.70}$\pm$0.94 & 29.54$\pm$12.2 & 67.99 \\
& \hspace{-0.2cm} \texttt{v19UNet} \hspace{-0.2cm} & 97.60$\pm$0.44 & 1.38$\pm$1.41 & 0.94$\pm$0.35 & 20.69$\pm$9.00 & 82.34 & 92.10$\pm$4.49 & 6.04$\pm$7.44 & 2.65$\pm$1.97 & 37.96$\pm$18.8 & 61.55 \\
& \hspace{-0.2cm} \texttt{v19pUNet} \hspace{-0.2cm} & 97.88$\pm$0.37 & 1.22$\pm$0.82 & 0.82$\pm$0.26 & 19.87$\pm$8.86 & 83.71 & 93.44$\pm$4.11 & 5.24$\pm$5.87 & 1.97$\pm$1.39 & 32.41$\pm$13.6 & 65.32 \\
& \hspace{-0.2cm} \texttt{v19UNet$+$} \cite{zhou2018unet} \hspace{-0.2cm} & 97.38$\pm$0.61 & 2.06$\pm$1.90 & 1.16$\pm$0.45 & 26.26$\pm$11.9 & 76.61 & 92.22$\pm$4.46 & 6.61$\pm$7.28 & 2.41$\pm$1.51 & 35.62$\pm$17.0 & 61.39 \\
& \hspace{-0.2cm} \texttt{v19pUNet$+$} \cite{zhou2018unet} \hspace{-0.2cm} & 97.80$\pm$0.42 & 1.49$\pm$1.45 & 0.85$\pm$0.26 & 18.97$\pm$6.71 & 82.69 & 92.83$\pm$6.92 & 5.91$\pm$8.73 & 2.12$\pm$2.04 & 31.54$\pm$18.3 & 66.14 \\
& \hspace{-0.2cm} \texttt{UNet\tiny{1-1}} \cite{yan19cascaded} \hspace{-0.2cm} & 97.48$\pm$0.61 & 1.64$\pm$1.82 & 1.02$\pm$0.59 & 20.89$\pm$10.9 & 81.28 & 92.03$\pm$4.04 & 5.81$\pm$6.73 & 2.45$\pm$1.43 & 34.04$\pm$15.9 & 63.05 \\
& \hspace{-0.2cm} \texttt{v16pUNet\tiny{1-1}} \hspace{-0.2cm} & 97.94$\pm$0.32 & \underline{1.12}$\pm$0.91 & \textbf{0.76}$\pm$0.16 & \textbf{17.08}$\pm$5.80 & \textbf{85.53} & 94.28$\pm$1.99 & 4.09$\pm$3.07 & 1.67$\pm$0.94 & \underline{28.60}$\pm$12.4 & 68.92 \\
& \hspace{-0.2cm} \texttt{v19pUNet\tiny{1-1}} \hspace{-0.2cm} & \underline{97.91}$\pm$0.26 & 1.14$\pm$0.95 & \underline{0.78}$\pm$0.17 & 19.44$\pm$7.46 & 84.40 & \textbf{94.52}$\pm$1.64 & \underline{3.52}$\pm$2.32 & \textbf{1.62}$\pm$1.02 & \textbf{27.02}$\pm$14.5 & \textbf{70.05} \\
& \hspace{-0.2cm} \texttt{cGv16pUNet\tiny{1-1}} \hspace{-0.2cm} & \textbf{97.95}$\pm$0.27 & 1.19$\pm$0.89 & \textbf{0.76}$\pm$0.16 & \underline{18.69}$\pm$7.58 & \underline{84.50} & 94.02$\pm$2.42 & 4.41$\pm$3.73 & 1.79$\pm$1.06 & 30.02$\pm$13.6 & 67.88 \\
& \hspace{-0.2cm} \texttt{cGv19pUNet\tiny{1-1}} \hspace{-0.2cm} & 97.87$\pm$0.32 & \textbf{1.09}$\pm$0.96 & 0.80$\pm$0.23 & 20.52$\pm$8.24 & 84.15 & \underline{94.33}$\pm$1.75 & \textbf{3.49}$\pm$2.57 & 1.73$\pm$0.97 & 28.94$\pm$15.0 & \underline{69.21} \\ \cline{2-12}
& \hspace{-0.2cm} \texttt{MOvpUNet} \hspace{-0.2cm} & 97.94$\pm$0.32 & 1.12$\pm$0.91 & 0.76$\pm$0.16 & 17.08$\pm$5.80 & 85.53 & 94.45$\pm$1.74 & 3.45$\pm$2.45 & 1.67$\pm$1.01 & 27.46$\pm$14.5 & 70.17 \\ \hline
\end{tabular}
\caption{Quantitative assessment of \texttt{DeepMedic} \cite{kamnitsas2017efficient}, \texttt{VNet} \cite{milletari2016vnet}, \texttt{denseVNet} \cite{gibson2018automatic}, \texttt{UNet} \cite{ronneberger2015unet}, \texttt{v16UNet}, \texttt{v16pUNet} \cite{conze19isbi}, \texttt{v19UNet}, \texttt{v19pUNet}, \texttt{v19UNet$+$} \cite{zhou2018unet}, \texttt{v19pUNet$+$} \cite{zhou2018unet}, \texttt{UNet\tiny{1-1}} \cite{yan19cascaded} and proposed \texttt{v16pUNet\tiny{1-1}},  \texttt{v19pUNet\tiny{1-1}}, \texttt{cGv16pUNet\tiny{1-1}}, \texttt{cGv19pUNet\tiny{1-1}} and \texttt{MOvpUNet} architectures for healthy liver segmentation in CT and MR images. Bold and underline results indicate first and second best scores.} \vspace{-0.3cm}
\label{tab::tab-1} \vspace{-0.2cm}
\end{table*}

\footnotetext[3]{CHAOS data available at \url{https://doi.org/10.5281/zenodo.3362844}}

Except for DeepMedic \cite{kamnitsas2017efficient}, VNet \cite{milletari2016vnet} and denseVNet \cite{gibson2018automatic}, evaluated models independently process 2D axial slices and produce 2D segmentation masks which are then stacked together to recover 3D volumes. Image size for axial slices are $256\hspace{-0.02cm}\times\hspace{-0.02cm}256$ or $288\hspace{-0.02cm}\times\hspace{-0.02cm}288$ pixels for MR images, $512\hspace{-0.02cm}\times\hspace{-0.02cm}512$ for CT examinations. The number of axial slices varies from $26$ to $50$ (resp. $78$ to $294$) and slice thickness is between 4.4 and 8.0 (2.0 and 3.2) mm for MR (resp. CT) images.

Let $S$ and $G$ deal with segmentation results and groundtruth. To assess standard CED (DeepMedic \cite{kamnitsas2017efficient}, VNet \cite{milletari2016vnet}, denseVNet \cite{gibson2018automatic}, \texttt{UNet} \cite{ronneberger2015unet}), deeper CED with or without encoder pre-training (\texttt{v16UNet}, \texttt{v16pUNet} \cite{conze19isbi}, \texttt{v19UNet}, \texttt{v19pUNet}), CED using nested and dense skip connections (\texttt{v19UNet$+$}, \texttt{v19pUNet$+$} \cite{zhou2018unet}), cascaded CED (\texttt{UNet\tiny{1-1}} \cite{yan19cascaded}, \texttt{v16pUNet\tiny{1-1}}, \texttt{v19pUNet\tiny{1-1}}) and cGAN with partially pre-trained cascaded CED as generator (\texttt{cGv16pUNet\tiny{1-1}}, \texttt{cGv19pUNet\tiny{1-1}}), the accuracy of abdominal organ segmentation is quantified based on Dice coefficient (\texttt{dice}) estimated following $\frac{2 |S \cap G|}{|S| + |G|}$ where $|.|$ denotes cardinality, relative absolute volume difference (\texttt{RAVD}) comparing $S$ and $G$ such as $\texttt{RAVD}=\frac{\texttt{abs}(|S|-|G|)}{|G|}$, average and maximum symmetric surface distances (\texttt{ASSD}, \texttt{MSSD}) which correspond to the average (maximum) Hausdorff distance between border voxels in $S$ and $G$. These metrics tend to provide an overall assessment of the involved networks. Following \cite{kavur2020chaos}, we also provide model ranking scores by averaging all metrics after having transformed values to span the $[0,100]$ interval so that higher values correspond to better segmentation. To discard unacceptable accuracy and increase metric sensitivity \cite{kavur2020chaos}, thresholds are set up according to intra/inter-expert similarities: $\texttt{dice}\hspace{-0.05cm}>\hspace{-0.05cm}80\%$, $\texttt{RAVD}\hspace{-0.05cm}<\hspace{-0.05cm}5\%$, $\texttt{ASSD}\hspace{-0.05cm}<\hspace{-0.05cm}15$mm and $\texttt{MSSD}\hspace{-0.05cm}<\hspace{-0.05cm}60$mm. Metrics outside the threshold range get zero points. Scores reached for multi-organ segmentation are obtained by averaging scores for each organ. Similarly, scores for MR images are the average between results arising from T1-DUAL\texttt{in/out} and T2-SPIR.

In our experiments, a given model is dedicated to one single modality (T1-DUAL\texttt{in/out}, T2-SPIR, CT) and one single organ (liver, right kidney, left kidney, spleen). Each model thus performs binary instead of multi-class segmentation to extract robust organ-specific features. In addition, experiments on the T1-DUAL modality stack together T1-DUAL\texttt{in} and T1-DUAL\texttt{out} images as model inputs since both phases are registered. When 3 channels are required, as for \texttt{v16}(\texttt{p})\texttt{UNet}, the third channel consists of the T1DUAL\texttt{in} image duplication. For CT and T2-SPIR, image content is replicated twice. 

Deep CEDs are trained using data augmentation to teach \textcolor{black}{networks} \textcolor{black}{efficient} invariance and robustness properties \cite{ronneberger2015unet}. Training axial slices undergo random scaling, rotation, shearing and shifting \textcolor{black}{operations}. 100 augmented images are produced for \textcolor{black}{a} single training slice. For CT (MR) images, models are trained with $6$ ($20$) epochs, a batch size of $3$ ($5$) images, an \textit{Adam} optimizer with $10^{-5}$ as learning rate and a fuzzy Dice score as loss function. Models were implemented using Keras and trained using Nvidia GeForce GTX 1080 Ti GPU. \vspace{-0.3cm}

\subsection{Evaluation on clinical data}
\label{sec:ssec4-2}

\noindent \textbf{CT and MR liver segmentation.} Quantitative metric and score values are provided in Tab.\ref{tab::tab-1} for liver CT/MR delineation. For both modalities, standard architectures including DeepMedic, VNet, denseVNet and \texttt{UNet} are outperformed by deeper (\texttt{v16}/\texttt{v16pUNet}, \texttt{v19}/\texttt{v19pUNet}) and cascaded (\texttt{UNet\tiny{1-1}}) networks which indicates that better predictive performance and generalizability is reached \textit{via} more complex models. In one hand, comparisons between \texttt{v16}/\texttt{v19UNet} and their partially pre-trained extensions (\texttt{v16p}/\texttt{v19pUNet}) reveals that pre-training the encoder using non-medical ImageNet data makes the network converge towards a better solution. In particular, large gains in terms of \texttt{dice} ($91.60$ to $94.07\%$) and \texttt{ASSD} ($2.96$ to $1.70$mm) are reported for MR \textcolor{black}{images} between \texttt{v16UNet} and \texttt{v16pUNet}. In the other hand, extending \texttt{UNet} into a cascaded pipeline (\texttt{UNet\tiny{1-1}}) through auto-context and end-to-end-training allows \textcolor{black}{taking} advantage of multi-level context, with score improvements from $79.07$ ($58.02$) to $81.28\%$ ($63.05\%$) \textcolor{black}{in} CT (MR). \texttt{v19UNet} and \texttt{v19pUNet} give better or slightly similar performance than their nested and dense counterparts (\texttt{v19}/\texttt{v19pUNet$+$}) which suggests that the great complexity brought by such heavy architectures is not useful to provide relevant liver contours.

\begin{table*}
\scriptsize 
\hspace{-0.1cm} \begin{tabular}{|c|l|ccccc|ccccc|}
\hline
\multirow{2}{*}{organ} & \multirow{2}{*}{\hspace{-0.1cm}model} & \multicolumn{5}{c|}{T1-DUAL\texttt{in/out}} & \multicolumn{5}{c|}{T2-SPIR} \\ \cline{3-12}
& & \texttt{dice} \textcolor{black}{$\uparrow$} & \texttt{RAVD} \textcolor{black}{$\downarrow$} & \texttt{ASSD} \textcolor{black}{$\downarrow$} & \texttt{MSSD} \textcolor{black}{$\downarrow$} & \textbf{score} & \texttt{dice} \textcolor{black}{$\uparrow$} & \texttt{RAVD} \textcolor{black}{$\downarrow$} & \texttt{ASSD} \textcolor{black}{$\downarrow$} & \texttt{MSSD} \textcolor{black}{$\downarrow$} & \textbf{score} \\ \hline
\multirow{15}{*}{\rotatebox{90}{liver}} & \hspace{-0.2cm} \texttt{DeepMedic} \cite{kamnitsas2017efficient} \hspace{-0.2cm} & 89.24$\pm$9.81 & 7.11$\pm$10.1 & 5.05$\pm$5.88 & 116.5$\pm$47.6 & 47.59 & 90.24$\pm$5.27 & 5.94$\pm$6.49 & 4.43$\pm$3.78 & 128.5$\pm$58.8 & 47.69 \\
& \hspace{-0.2cm} \texttt{denseVNet} \cite{gibson2018automatic} \hspace{-0.2cm} & 85.56$\pm$6.10 & 17.2$\pm$12.3 & 4.63$\pm$2.67 & 43.10$\pm$23.6 & 45.55 & 67.94$\pm$7.62 & 17.6$\pm$13.8 & 11.9$\pm$3.58 & 66.86$\pm$33.3 & 12.26 \\
& \hspace{-0.2cm} \texttt{UNet} \cite{ronneberger2015unet} \hspace{-0.2cm} & 90.48$\pm$7.44 & 9.57$\pm$12.4 & 2.74$\pm$1.99 & 35.39$\pm$19.1 & 60.85 & 90.52$\pm$3.34 & 6.44$\pm$5.14 & 3.84$\pm$2.79 & 53.59$\pm$29.0 & 55.14 \\
& \hspace{-0.2cm} \texttt{v16UNet} \hspace{-0.2cm} & 91.32$\pm$7.66 & 8.38$\pm$12.3 & 2.36$\pm$1.94 & 29.62$\pm$18.0 & 63.43 & 91.75$\pm$3.19 & 5.54$\pm$4.94 & 3.61$\pm$2.80 & 52.42$\pm$32.6 & 57.63 \\
& \hspace{-0.2cm} \texttt{v16pUNet} \cite{conze19isbi} \hspace{-0.2cm} & 93.64$\pm$2.84 & 5.77$\pm$5.28 & 1.79$\pm$0.92 & 31.17$\pm$11.7 & 64.81 & 94.46$\pm$1.82 & 2.79$\pm$1.73 & 1.65$\pm$0.96 & 28.97$\pm$12.7 & 70.56 \\
& \hspace{-0.2cm} \texttt{v19UNet} \hspace{-0.2cm} & 91.83$\pm$6.28 & 7.38$\pm$10.3 & 2.32$\pm$1.75 & 32.04$\pm$16.7 & 62.29 & 92.40$\pm$2.70 & 4.82$\pm$4.49 & 2.94$\pm$2.21 & 43.86$\pm$20.9 & 60.21 \\
& \hspace{-0.2cm} \texttt{v19pUNet} \hspace{-0.2cm} & 92.39$\pm$6.30 & 7.71$\pm$9.96 & 2.31$\pm$1.81 & 35.70$\pm$14.3 & 59.77 & 94.48$\pm$1.92 & 2.84$\pm$1.80 & 1.63$\pm$0.96 & 29.17$\pm$12.8 & 70.56 \\
& \hspace{-0.2cm} \texttt{v19UNet$+$} \cite{zhou2018unet} \hspace{-0.2cm} & 91.64$\pm$6.01 & 7.94$\pm$10.2 & 2.44$\pm$1.58 & 36.62$\pm$19.8 & 60.24 & 92.68$\pm$2.89 & 5.25$\pm$4.51 & 2.39$\pm$1.44 & 34.61$\pm$14.1 & 62.91 \\
& \hspace{-0.2cm} \texttt{v19pUNet$+$} \cite{zhou2018unet} \hspace{-0.2cm} & 91.17$\pm$12.1 & 8.86$\pm$15.7 & 2.54$\pm$3.13 & 31.47$\pm$19.7 & 63.23 & 94.48$\pm$1.69 & 3.03$\pm$1.81 & 1.69$\pm$0.95 & 31.62$\pm$16.9 & 68.71 \\
& \hspace{-0.2cm} \texttt{UNet\tiny{1-1}} \cite{yan19cascaded} \hspace{-0.2cm} & 92.06$\pm$5.34 & 6.35$\pm$9.10 & 2.30$\pm$1.61 & 31.45$\pm$16.4 & 65.22 & 91.97$\pm$2.74 & 5.41$\pm$4.43 & 2.56$\pm$1.24 & 36.42$\pm$14.9 & 60.74 \\
& \hspace{-0.2cm} \texttt{v16pUNet\tiny{1-1}} \hspace{-0.2cm} & 93.86$\pm$2.28 & 5.49$\pm$4.72 & \underline{1.77}$\pm$0.88 & 29.87$\pm$10.6 & 65.93 & 93.69$\pm$1.68 & \textbf{2.66}$\pm$1.41 & \textbf{1.57}$\pm$0.99 & \underline{27.31}$\pm$13.8 & \underline{72.04} \\
& \hspace{-0.2cm} \texttt{v19pUNet\tiny{1-1}} \hspace{-0.2cm} & \textbf{94.38}$\pm$1.33 & \underline{4.25}$\pm$3.17 & 1.68$\pm$0.97 & \textbf{28.45}$\pm$14.9 & \underline{67.70} & \textbf{94.67}$\pm$1.92 & 2.80$\pm$1.49 & \underline{1.57}$\pm$1.06 & \textbf{25.49}$\pm$14.1 & \textbf{72.28} \\
& \hspace{-0.2cm} \texttt{cGv16pUNet\tiny{1-1}} \hspace{-0.2cm} & 93.45$\pm$2.96 & 6.05$\pm$5.77 & 1.96$\pm$1.05 & 32.30$\pm$13.4 & 64.57 & \underline{94.60}$\pm$1.89 & \underline{2.78}$\pm$1.77 & 1.63$\pm$1.07 & 27.74$\pm$13.9 & 71.26 \\
& \hspace{-0.2cm} \texttt{cGv19pUNet\tiny{1-1}} \hspace{-0.2cm} & \underline{94.23}$\pm$1.54 & \textbf{4.17}$\pm$3.42 & \textbf{1.76}$\pm$0.96 & \underline{29.42}$\pm$15.0 & \textbf{67.72} & 94.44$\pm$1.95 & 2.86$\pm$1.71 & 1.71$\pm$1.00 & 28.46$\pm$15.0 & 70.49 \\ \hline
\multirow{15}{*}{\rotatebox{90}{right kidney}} & \hspace{-0.2cm} \texttt{DeepMedic} \cite{kamnitsas2017efficient} \hspace{-0.2cm} & 75.13$\pm$15.8 & 29.06$\pm$19.8 & 3.73$\pm$2.11 & 105.8$\pm$62.5 & 35.42 & 89.32$\pm$10.0 & 11.83$\pm$14.7 & 1.66$\pm$1.28 & 102.5$\pm$63.9 & 51.40 \\
& \hspace{-0.2cm} \texttt{denseVNet} \cite{gibson2018automatic} \hspace{-0.2cm} & 76.31$\pm$11.2 & 25.66$\pm$20.0 & 4.39$\pm$2.43 & 34.72$\pm$38.1 & 41.63 & 67.94$\pm$6.92 & 21.00$\pm$14.1 & 7.59$\pm$4.00 & 61.14$\pm$47.2 & 23.61 \\
& \hspace{-0.2cm} \texttt{UNet} \cite{ronneberger2015unet} \hspace{-0.2cm} & 85.61$\pm$13.2 & 13.44$\pm$16.9 & 2.55$\pm$3.45 & 20.26$\pm$15.6 & 61.05 & 88.16$\pm$5.77 & 15.39$\pm$15.2 & 3.70$\pm$4.54 & 28.85$\pm$22.8 & 55.87 \\
& \hspace{-0.2cm} \texttt{v16UNet} \hspace{-0.2cm} & 87.19$\pm$6.11 & 11.52$\pm$10.5 & 2.61$\pm$2.82 & 23.83$\pm$16.3 & 57.85 & 91.68$\pm$3.72 & 9.85$\pm$6.54 & 1.42$\pm$1.14 & 18.00$\pm$10.4 & 64.19 \\
& \hspace{-0.2cm} \texttt{v16pUNet} \cite{conze19isbi} \hspace{-0.2cm} & 90.08$\pm$3.88 & 11.55$\pm$7.40 & 1.38$\pm$0.86 & 12.35$\pm$7.22 & 66.32 & 92.47$\pm$3.96 & 8.56$\pm$5.04 & 1.08$\pm$1.12 & \underline{12.64}$\pm$8.19 & 67.79 \\
& \hspace{-0.2cm} \texttt{v19UNet} \hspace{-0.2cm} & 87.36$\pm$7.39 & 13.67$\pm$14.7 & 2.12$\pm$2.03 & 20.27$\pm$14.6 & 61.69 & 92.12$\pm$3.91 & 8.89$\pm$6.16 & 1.28$\pm$1.19 & 15.30$\pm$8.80 & 66.59 \\
& \hspace{-0.2cm} \texttt{v19pUNet} \hspace{-0.2cm} & 90.26$\pm$4.28 & \underline{10.96}$\pm$8.39 & 1.29$\pm$0.81 & \textbf{11.58}$\pm$6.35 & \textbf{68.07} & 92.66$\pm$4.08 & \underline{7.85}$\pm$5.11 & 1.07$\pm$1.14 & 12.65$\pm$8.77 & 68.38 \\
& \hspace{-0.2cm} \texttt{v19UNet$+$} \cite{zhou2018unet} \hspace{-0.2cm} & 86.49$\pm$9.47 & 14.40$\pm$13.1 & 2.09$\pm$1.31 & 19.32$\pm$9.03 & 59.67 & 86.43$\pm$20.8 & 10.18$\pm$7.28 & 6.61$\pm$22.8 & 24.02$\pm$29.6 & 61.91 \\
& \hspace{-0.2cm} \texttt{v19pUNet$+$} \cite{zhou2018unet} \hspace{-0.2cm} & 89.34$\pm$5.48 & 13.75$\pm$8.42 & 1.46$\pm$0.99 & 12.90$\pm$6.74 & 64.50 & 92.82$\pm$3.33 & 8.46$\pm$4.39 & 1.16$\pm$1.14 & 15.32$\pm$9.95 & 67.56 \\
& \hspace{-0.2cm} \texttt{UNet\tiny{1-1}} \cite{yan19cascaded} \hspace{-0.2cm} & 86.32$\pm$9.30 & 14.82$\pm$14.9 & 2.11$\pm$1.71 & 17.15$\pm$10.3 & 60.61 & 91.17$\pm$4.46 & 9.04$\pm$8.30 & 1.77$\pm$1.90 & 18.07$\pm$13.6 & 65.75 \\
& \hspace{-0.2cm} \texttt{v16pUNet\tiny{1-1}} \hspace{-0.2cm} & 90.27$\pm$3.19 & 11.92$\pm$6.32 & \textbf{1.32}$\pm$0.72 & \underline{11.95}$\pm$7.25 & 66.22 & 92.78$\pm$4.19 & 8.71$\pm$5.32 & 1.02$\pm$1.14 & 12.46$\pm$7.87 & 68.39 \\
& \hspace{-0.2cm} \texttt{v19pUNet\tiny{1-1}} \hspace{-0.2cm} & \underline{90.30}$\pm$3.73 & 11.66$\pm$7.28 & 1.47$\pm$1.16 & 13.96$\pm$10.5 & 66.19 & \underline{93.21}$\pm$2.84 & \textbf{7.76}$\pm$6.26 & \textbf{0.97}$\pm$1.05 & 12.63$\pm$8.12 & \textbf{72.71} \\
& \hspace{-0.2cm} \texttt{cGv16pUNet\tiny{1-1}} \hspace{-0.2cm} & 90.29$\pm$3.91 & 11.38$\pm$8.50 & 1.38$\pm$1.02 & 13.60$\pm$8.18 & \underline{67.26} & \textbf{93.22}$\pm$3.45 & 8.06$\pm$7.89 & 1.01$\pm$1.06 & 15.87$\pm$10.2 & \underline{71.34} \\
& \hspace{-0.2cm} \texttt{cGv19pUNet\tiny{1-1}} \hspace{-0.2cm} & \textbf{90.56}$\pm$4.28 & \textbf{10.44}$\pm$8.92 & \underline{1.37}$\pm$0.99 & 13.39$\pm$9.21 & 66.67 & 93.02$\pm$3.74 & 7.94$\pm$6.77 & \underline{0.99}$\pm$1.09 & \textbf{11.23}$\pm$7.80 & 71.21 \\ \hline
\multirow{15}{*}{\rotatebox{90}{left kidney}} & \hspace{-0.2cm} \texttt{DeepMedic} \cite{kamnitsas2017efficient} \hspace{-0.2cm} & 69.95$\pm$21.7 & 34.17$\pm$24.4 & 5.81$\pm$6.01 & 123.7$\pm$56.8 & 28.38 & 80.36$\pm$23.8 & 20.43$\pm$27.1 & 3.52$\pm$3.15 & 120.1$\pm$70.4 & 45.48 \\
& \hspace{-0.2cm} \texttt{denseVNet} \cite{gibson2018automatic} \hspace{-0.2cm} & 68.71$\pm$22.7 & 25.83$\pm$26.7 & 32.1$\pm$120. & 61.22$\pm$119. & 40.12 & 64.84$\pm$11.8 & 23.01$\pm$19.3 & 7.81$\pm$3.64 & 48.08$\pm$36.9 & 24.02 \\
& \hspace{-0.2cm} \texttt{UNet} \cite{ronneberger2015unet} \hspace{-0.2cm} & 81.55$\pm$16.8 & 16.86$\pm$18.5 & 4.25$\pm$6.99 & 37.02$\pm$30.3 & 51.71 & 90.32$\pm$3.47 & 9.34$\pm$6.70 & 2.11$\pm$1.37 & 36.19$\pm$23.8 & 58.95 \\
& \hspace{-0.2cm} \texttt{v16UNet} \hspace{-0.2cm} & 83.33$\pm$15.3 & 16.94$\pm$17.6 & 3.04$\pm$3.65 & \underline{30.92}$\pm$31.5 & 53.49 & 91.52$\pm$2.52 & 9.32$\pm$5.06 & 1.74$\pm$1.41 & 26.91$\pm$24.5 & 62.75 \\
& \hspace{-0.2cm} \texttt{v16pUNet} \cite{conze19isbi} \hspace{-0.2cm} & \underline{85.79}$\pm$20.4 & \textbf{10.13}$\pm$6.37 & 8.91$\pm$33.7 & 34.53$\pm$50.2 & \textbf{63.52} & \textbf{92.83}$\pm$2.14 & \underline{8.18}$\pm$5.20 & \textbf{1.32}$\pm$1.27 & 24.08$\pm$22.3 & 64.58 \\
& \hspace{-0.2cm} \texttt{v19UNet} \hspace{-0.2cm} & 82.06$\pm$20.3 & 16.45$\pm$16.4 & 7.97$\pm$23.3 & 44.09$\pm$55.9 & 52.73 & 90.64$\pm$4.01 & 9.90$\pm$5.45 & 1.77$\pm$1.55 & 26.94$\pm$22.9 & 61.76 \\
& \hspace{-0.2cm} \texttt{v19pUNet} \hspace{-0.2cm} & 85.58$\pm$20.5 & 14.38$\pm$16.5 & 8.68$\pm$31.5 & 34.84$\pm$51.1 & 59.64 & 92.60$\pm$2.39 & 9.00$\pm$6.11 & 1.57$\pm$2.08 & 23.62$\pm$22.2 & 63.52 \\
& \hspace{-0.2cm} \texttt{v19UNet$+$} \cite{zhou2018unet} \hspace{-0.2cm} & 82.69$\pm$20.4 & 16.49$\pm$19.8 & 7.72$\pm$23.9 & 39.19$\pm$57.6 & 56.03 & 90.97$\pm$4.05 & 11.59$\pm$10.1 & 2.26$\pm$2.59 & 31.25$\pm$24.9 & 58.61 \\
& \hspace{-0.2cm} \texttt{v19pUNet$+$} \cite{zhou2018unet} \hspace{-0.2cm} & \textbf{88.76}$\pm$7.87 & 13.94$\pm$10.3 & \textbf{1.62}$\pm$1.21 & \textbf{27.17}$\pm$30.3 & 61.03 & 92.79$\pm$3.00 & 9.01$\pm$6.06 & 1.49$\pm$2.10 & \underline{22.69}$\pm$23.0 & 64.54 \\
& \hspace{-0.2cm} \texttt{UNet\tiny{1-1}} \cite{yan19cascaded} \hspace{-0.2cm} & 83.88$\pm$12.2 & 16.72$\pm$16.2 & \underline{2.91}$\pm$2.33 & 33.55$\pm$29.7 & 51.35 & 89.91$\pm$4.65 & 10.89$\pm$7.62 & 1.67$\pm$1.03 & 23.50$\pm$20.4 & 64.30 \\
& \hspace{-0.2cm} \texttt{v16pUNet\tiny{1-1}} \hspace{-0.2cm} & 85.56$\pm$20.5 & \underline{11.40}$\pm$9.22 & 8.58$\pm$31.7 & 37.23$\pm$48.2 & \underline{61.66} & 92.78$\pm$2.97 & 8.76$\pm$7.59 & 1.46$\pm$2.11 & \textbf{22.04}$\pm$23.2 & 64.57 \\
& \hspace{-0.2cm} \texttt{v19pUNet\tiny{1-1}} \hspace{-0.2cm} & 84.01$\pm$20.5 & 14.18$\pm$10.1 & 9.01$\pm$32.2 & 35.68$\pm$49.2 & 56.64 & 92.10$\pm$3.03 & 9.45$\pm$8.11 & 1.87$\pm$2.79 & 24.31$\pm$23.0 & 62.63 \\
& \hspace{-0.2cm} \texttt{cGv16pUNet\tiny{1-1}} \hspace{-0.2cm} & 84.70$\pm$20.4 & 12.10$\pm$9.23 & 8.70$\pm$30.3 & 38.39$\pm$48.6 & 56.13 & \underline{92.83}$\pm$2.27 & \textbf{8.05}$\pm$5.40 & \underline{1.35}$\pm$1.41 & 23.90$\pm$22.3 & \textbf{65.56} \\
& \hspace{-0.2cm} \texttt{cGv19pUNet\tiny{1-1}} \hspace{-0.2cm} & 85.31$\pm$20.4 & 13.17$\pm$15.1 & 7.04$\pm$23.8 & 36.08$\pm$49.1 & 59.89 & 92.67$\pm$3.30 & 8.88$\pm$8.80 & 1.67$\pm$2.91 & 23.89$\pm$23.6 & \underline{64.64} \\ \hline
\multirow{15}{*}{\rotatebox{90}{spleen}} & \hspace{-0.2cm} \texttt{DeepMedic} \cite{kamnitsas2017efficient} \hspace{-0.2cm} & 75.33$\pm$24.2 & 23.23$\pm$27.4 & 5.59$\pm$6.74 & 155.3$\pm$66.2 & 35.16 & 88.24$\pm$5.57 & 13.15$\pm$8.59 & 5.07$\pm$4.75 & 165.8$\pm$87.9 & 40.50 \\
& \hspace{-0.2cm} \texttt{denseVNet} \cite{gibson2018automatic} \hspace{-0.2cm} & 69.38$\pm$17.3 & 31.26$\pm$19.8 & 6.18$\pm$3.98 & 61.09$\pm$71.7 & 31.06 & 48.56$\pm$19.3 & 25.11$\pm$17.0 & 16.1$\pm$9.94 & 90.46$\pm$67.5 & 9.33 \\
& \hspace{-0.2cm} \texttt{UNet} \cite{ronneberger2015unet} \hspace{-0.2cm} & 81.56$\pm$19.8 & 20.68$\pm$25.0 & 3.38$\pm$4.08 & 26.41$\pm$18.9 & 53.99 & 89.33$\pm$5.83 & 8.56$\pm$7.73 & 2.02$\pm$1.86 & 23.46$\pm$16.2 & 59.85 \\
& \hspace{-0.2cm} \texttt{v16UNet} \hspace{-0.2cm} & 85.20$\pm$9.90 & 17.67$\pm$17.9 & 2.71$\pm$2.57 & 29.08$\pm$22.4 & 55.51 & 89.84$\pm$6.51 & 10.14$\pm$8.97 & 2.04$\pm$2.32 & 22.01$\pm$17.4 & 62.22 \\
& \hspace{-0.2cm} \texttt{v16pUNet} \cite{conze19isbi} \hspace{-0.2cm} & 89.66$\pm$3.75 & 11.79$\pm$7.30 & \underline{1.41}$\pm$0.77 & 15.24$\pm$11.5 & 66.35 & 92.60$\pm$3.29 & 7.92$\pm$5.43 & 1.03$\pm$0.63 & 15.04$\pm$9.68 & 68.93 \\
& \hspace{-0.2cm} \texttt{v19UNet} \hspace{-0.2cm} & 82.40$\pm$20.6 & 16.51$\pm$14.0 & 6.80$\pm$20.3 & 31.72$\pm$35.5 & 56.09 & 89.74$\pm$5.46 & 9.63$\pm$7.99 & 1.97$\pm$1.65 & 23.22$\pm$13.8 & 60.90 \\
& \hspace{-0.2cm} \texttt{v19pUNet} \hspace{-0.2cm} & 89.59$\pm$3.88 & \underline{11.60}$\pm$8.06 & 1.57$\pm$0.97 & 17.87$\pm$15.1 & 66.28 & 92.17$\pm$3.71 & 9.27$\pm$5.48 & 1.10$\pm$0.75 & 16.86$\pm$10.7 & 66.82 \\
& \hspace{-0.2cm} \texttt{v19UNet$+$} \cite{zhou2018unet} \hspace{-0.2cm} & 84.24$\pm$12.6 & 18.48$\pm$18.4 & 2.76$\pm$2.73 & 23.94$\pm$17.9 & 56.75 & 89.34$\pm$6.76 & 10.87$\pm$9.70 & 1.96$\pm$2.24 & 21.49$\pm$12.2 & 62.05 \\
& \hspace{-0.2cm} \texttt{v19pUNet$+$} \cite{zhou2018unet} \hspace{-0.2cm} & 89.54$\pm$4.10 & 11.81$\pm$8.89 & 1.43$\pm$0.88 & \textbf{13.90}$\pm$9.25 & \textbf{69.04} & 92.14$\pm$4.48 & 7.82$\pm$6.43 & 1.29$\pm$1.51 & 16.40$\pm$13.5 & 68.56 \\
& \hspace{-0.2cm} \texttt{UNet\tiny{1-1}} \cite{yan19cascaded} \hspace{-0.2cm} & 87.01$\pm$7.34 & 12.03$\pm$11.0 & 1.98$\pm$1.31 & 24.38$\pm$13.6 & 61.43 & 86.60$\pm$10.5 & 14.10$\pm$15.9 & 3.25$\pm$4.51 & 25.01$\pm$22.0 & 58.58 \\
& \hspace{-0.2cm} \texttt{v16pUNet\tiny{1-1}} \hspace{-0.2cm} & 88.89$\pm$4.64 & 11.80$\pm$6.41 & 1.51$\pm$0.78 & 19.24$\pm$11.8 & 62.86 & \textbf{93.16}$\pm$3.26 & \textbf{7.57}$\pm$4.63 & \underline{0.86}$\pm$0.57 & 12.15$\pm$7.89 & 69.83 \\
& \hspace{-0.2cm} \texttt{v19pUNet\tiny{1-1}} \hspace{-0.2cm} & \textbf{89.93}$\pm$3.79 & 11.61$\pm$7.07 & \textbf{1.34}$\pm$0.66 & \underline{14.89}$\pm$9.21 & 66.94 & 92.29$\pm$4.03 & 8.68$\pm$6.74 & 1.13$\pm$0.94 & 15.37$\pm$11.0 & 68.64 \\
& \hspace{-0.2cm} \texttt{cGv16pUNet\tiny{1-1}} \hspace{-0.2cm} & \underline{89.67}$\pm$3.92 & \textbf{10.90}$\pm$7.78 & 1.45$\pm$0.78 & 15.82$\pm$9.34 & \underline{67.02} & \underline{93.00}$\pm$3.03 & \underline{7.70}$\pm$5.29 & \textbf{0.84}$\pm$0.44 & \textbf{10.96}$\pm$3.58 & \textbf{70.79} \\
& \hspace{-0.2cm} \texttt{cGv19pUNet\tiny{1-1}} \hspace{-0.2cm} & 89.31$\pm$3.88 & 11.85$\pm$6.25 & 1.63$\pm$1.08 & 19.21$\pm$15.4 & 63.79 & 92.41$\pm$3.58 & 9.17$\pm$5.73 & 1.05$\pm$0.91 & \underline{11.61}$\pm$7.85 & \underline{70.45} \\ \hline
\end{tabular} \vspace{-0.05cm}
\caption{Quantitative assessment of \texttt{DeepMedic} \cite{kamnitsas2017efficient}, \texttt{denseVNet} \cite{gibson2018automatic}, \texttt{UNet} \cite{ronneberger2015unet}, \texttt{v16UNet}, \texttt{v16pUNet} \cite{conze19isbi}, \texttt{v19UNet}, \texttt{v19pUNet}, \texttt{v19UNet$+$} \cite{zhou2018unet}, \texttt{v19pUNet$+$} \cite{zhou2018unet}, \texttt{UNet\tiny{1-1}} \cite{yan19cascaded} and proposed \texttt{v16pUNet\tiny{1-1}},  \texttt{v19pUNet\tiny{1-1}}, \texttt{cGv16pUNet\tiny{1-1}} and \texttt{cGv19pUNet\tiny{1-1}} architectures for healthy abdominal multi-organ (liver, \textcolor{black}{right kidney}, \textcolor{black}{left kidney}, spleen) segmentation in T1-DUAL\texttt{in/out} and T2-SPIR images. Bold and underline results indicate first and second best scores.} \vspace{-0.4cm} 
\label{tab::tab-2}
\end{table*}

\begin{figure*}
\scriptsize
\hspace{-0.2cm} \begin{minipage}{0.5\linewidth}
\begin{tabular}{cccccc}
\rotatebox{90}{\textcolor{white}{------} T1-DUAL\texttt{in}} &
\hspace{-0.3cm} \includegraphics[height=2.63cm]{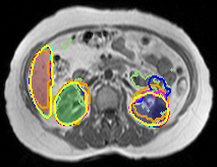} &
\hspace{-0.4cm} \includegraphics[height=2.63cm]{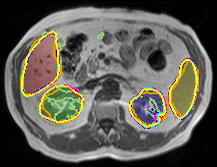} &
\hspace{-0.4cm} \includegraphics[height=2.63cm]{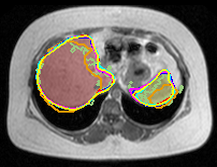} &
\hspace{-0.4cm} \includegraphics[height=2.63cm]{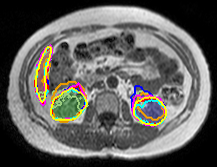} &
\hspace{-0.4cm} \includegraphics[height=2.63cm]{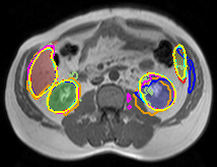} \cr
\rotatebox{90}{\textcolor{white}{----------} T2-SPIR} &
\hspace{-0.3cm} \includegraphics[height=2.63cm]{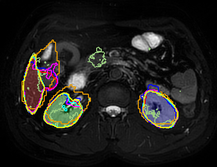} &
\hspace{-0.4cm} \includegraphics[height=2.63cm]{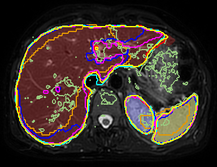} &
\hspace{-0.4cm} \includegraphics[height=2.63cm]{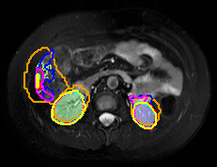} &
\hspace{-0.4cm} \includegraphics[height=2.63cm]{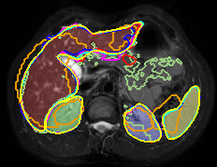} &
\hspace{-0.4cm} \includegraphics[height=2.63cm]{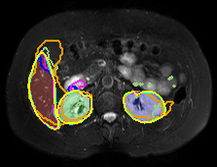} \cr
\end{tabular}
\end{minipage} \vspace{0.15cm} \\
\textcolor{color7}{---} \texttt{DeepMedic} \cite{kamnitsas2017efficient}
\textcolor{color8}{---} \texttt{denseVNet} \cite{gibson2018automatic}
\textcolor{color1}{---} \texttt{UNet} \cite{ronneberger2015unet}
\textcolor{color2}{---} \texttt{v19pUNet}
\textcolor{color3}{---} \texttt{v19pUNet$+$} \cite{zhou2018unet}
\textcolor{color4}{---} \texttt{v19pUNet\tiny{1-1}}
\textcolor{color5}{---} \texttt{cGv19pUNet\tiny{1-1}}
\textcolor{color10}{$\blacksquare$} liver
\textcolor{color11}{$\blacksquare$} r.kidney
\textcolor{color12}{$\blacksquare$} l.kidney
\textcolor{color13}{$\blacksquare$} spleen
\caption{Abdominal multi-organ MRI (T1-DUAL\texttt{in}, T2-SPIR) segmentation using \texttt{DeepMedic} \cite{kamnitsas2017efficient}, \texttt{denseVNet} \cite{gibson2018automatic}, \texttt{UNet} \cite{ronneberger2015unet}, \texttt{v19pUNet}, \texttt{v19pUNet$+$} \cite{zhou2018unet}, proposed \texttt{v19pUNet\tiny{1-1}} and \texttt{cGv19pUNet\tiny{1-1}}. Liver, right kidney, left kidney and spleen groundtruth delineations are respectively superimposed in red, green, blue and yelow colors.}
\label{fig::fig-8}
\end{figure*}

By combining these three contributions (deeper model, encoder pre-training and cascaded architecture), \texttt{v16pUNet\tiny{1-1}} (\texttt{v19pUNet\tiny{1-1}}) discriminates more efficiently liver areas from surrounding structures by achieving the best score for CT (MR) scans with $85.53\%$ ($70.05\%$). Embedding \texttt{v16pUNet\tiny{1-1}} (\texttt{v19pUNet\tiny{1-1}}) into a cGAN pipeline for CT (MR) liver segmentation gives broadly similar results but provides the second best scores. We note that \texttt{cGv16pUNet\tiny{1-1}} reaches the best \texttt{dice} ($97.95\%$) and similar \texttt{ASSD} ($0.76$mm) in CT. In MR, the best \texttt{RAVD} ($3.49\%$) is obtained using \texttt{cGv19pUNet\tiny{1-1}}.

Qualitative results for CT and MR liver segmentation are displayed in Fig. \ref{fig::fig-6}-\ref{fig::fig-7}. Compared to standard networks \textcolor{black}{as well as} \texttt{v16pUNet} (\texttt{v19pUNet}/\texttt{v19pUNet+}) which are prone to under- or over-segmentation, sometimes combined with unrealistic shapes, better contour adherence and shape consistency are reached by \texttt{v16pUNet\tiny{1-1}} (\texttt{v19pUNet\tiny{1-1}}) and \texttt{cGv16pUNet\tiny{1-1}} (\texttt{cGv19pUNet\tiny{1-1}}) whose ability to \textcolor{black}{mimic} expert annotations is notable for CT (T1-DUAL and T2-SPIR). The diversity in terms of textures arising in MR images is accurately captured through cascaded partially pre-trained networks despite high similar visual properties with surrounding structures. Moreover, deep networks find it harder to segment abdominal MR compared to CT images due to lower contrast and resolution combined with higher spacing. \\

\noindent \textbf{Abdominal multi-organ MR segmentation.} Tab.\ref{tab::tab-2} shows quantitative results for multi-organ MR segmentation. As for liver, DeepMedic, VNet, denseVNet, \texttt{UNet}, \texttt{v16UNet} and \texttt{v19UNet} networks do not provide the required robustness for organ extraction. In T1-DUAL modality, significant improvements can be noticed using \texttt{v16pUNet} for left kidney, \texttt{v19pUNet} for right kidney and \texttt{v19pUNet$+$} for spleen with the best reached scores ($63.52$, $68.07$ and $69.04\%$) among all schemes. Except for spleen in T2-SPIR, the comparison \texttt{UNet}/\texttt{UNet\tiny{1-1}} indicates the appropriateness of exploiting networks in a cascaded fashion, as proven for spleen (right kidney) in T1-DUAL (T2-SPIR) whose \texttt{dice} (\texttt{RAVD}) jumps from $81.56$ ($15.39$) to $87.01\%$ ($9.04\%$). The same conclusion arises between \texttt{v16p}/\texttt{v19pUNet} and \texttt{v16p}/\texttt{v19pUNet\tiny{1-1}} with\textcolor{black}{, for instance,} a strong score improvement \textcolor{black}{got using \texttt{v19pUNet\tiny{1-1}} for right kidney in T2-SPIR} ($68.38$ to $72.71\%$). Cascaded pre-trained cGAN strategies (\texttt{cGv16p}/\texttt{cGv19pUNet\tiny{1-1}}) always belong to one of the two best methods in \texttt{dice}, except for left kidney in T1-DUAL. Gains for \texttt{MSSD} in T2-SPIR are highlighted with $10.96$mm ($11.23$) \textcolor{black}{obtained} for spleen (right kidney) with \texttt{cGv16pUNet\tiny{1-1}} (\texttt{cGv19pUNet\tiny{1-1}}). Unsurprisingly, delineating small organs (kidneys) is more \textcolor{black}{challenging} than focusing on \textcolor{black}{larger ones (liver, spleen)}. The vicinity between left kidney and spleen further complicates the contouring. MR segmentation is easier with T2-SPIR than T1-DUAL since relative contrasts between structures is enhanced. 

\begin{table*}
\scriptsize 
\centering \begin{tabular}{|l|cl|cl|cl|cl|cl|cl|}
\hline
\multirow{3}{*}{model} & \multicolumn{2}{c|}{CT} & \multicolumn{10}{c|}{MRI} \\ \cline{2-13} 
& \multicolumn{2}{c|}{liver} & \multicolumn{2}{c|}{liver} & \multicolumn{2}{c|}{right kidney} & \multicolumn{2}{c|}{left kidney} & \multicolumn{2}{c|}{spleen} & \multicolumn{2}{c|}{multi-organ} \\ \cline{2-13} 
& \textbf{score} & rank & \textbf{score} & rank & \textbf{score} & rank & \textbf{score} & rank & \textbf{score} & rank & \textbf{score} & rank \\ \hline
\texttt{DeepMedic} \cite{kamnitsas2017efficient} \hspace{-0.2cm} & 73.32 & 14 & 47.64 & 13 & 43.41 & 13 & 36.93 & 13 & 37.83 & 13 & 41.45 & 13 \\
\texttt{denseVNet} \cite{gibson2018automatic} \hspace{-0.2cm} & 73.78 & 13 & 28.91 & 14 & 32.62 & 14 & 32.07 & 14 & 20.20 & 14 & 28.45 & 14 \\
\texttt{UNet} \cite{ronneberger2015unet} \hspace{-0.2cm} & 79.07 & 11 & 58.02 & 12 & 58.46 & 12 & 55.33 & 12 & 56.92 & 12 & 57.18 & 12 \\
\texttt{v16UNet} \hspace{-0.2cm} & 82.71 & 7 & 60.86 & 11 & 61.02 & 10 & 58.12 & 8 & 58.86 & 10 & 59.63 & 11 \\
\texttt{v16pUNet} \cite{conze19isbi} \hspace{-0.2cm} & 83.71 & 5/6 & 67.99 & 4 & 67.05 & 6 & \textbf{64.05} & \textbf{1} & 67.64 & 4 & 66.61 & 4 \\
\texttt{v19UNet} \hspace{-0.2cm} & 82.34 & 9 & 61.55 & 9 & 64.14 & 8 & 57.25 & 11 & 58.50 & 11 & 60.28 & 9 \\
\texttt{v19pUNet} \hspace{-0.2cm} & 83.71 & 5/6 & 65.32 & 7 & 68.23 & 4 & 61.58 & 5 & 66.55 & 6 & 65.38 & 7 \\
\texttt{v19UNet$+$} \cite{zhou2018unet} \hspace{-0.2cm} & 76.61 & 12 & 61.39 & 10 & 60.79 & 11 & 57.32 & 10 & 59.40 & 9 & 59.77 & 10 \\
\texttt{v19pUNet$+$} \cite{zhou2018unet} \hspace{-0.2cm} & 82.69 & 8 & 66.14 & 6 & 66.03 & 7 & \textit{62.79} & \textit{3} & \underline{68.80} & \underline{2} & 65.90 & 6 \\
\texttt{UNet\tiny{1-1}} \cite{yan19cascaded} \hspace{-0.2cm} & 81.28 & 10 & 63.05 & 8 & 63.18 & 9 & 57.82 & 9 & 60.01 & 8 & 61.00 & 8 \\
\texttt{v16pUNet\tiny{1-1}} \hspace{-0.2cm} & \textbf{85.53} & \textbf{1} & \textit{68.92} & \textit{3} & 67.31 & 5 & \underline{63.11} & \underline{2} & 66.35 & 7 & 66.44 & 5 \\
\texttt{v19pUNet\tiny{1-1}} \hspace{-0.2cm} & \textit{84.40} & \textit{3} & \textbf{70.05} & \textbf{1} & \textbf{69.45} & \textbf{1} & 59.64 & 7 & \textit{67.79} & \textit{3} & \textit{66.72} & \textit{3} \\
\texttt{cGv16pUNet\tiny{1-1}} \hspace{-0.2cm} & \underline{84.50} & \underline{2} & 67.88 & 5 & \underline{69.30} & \underline{2} & 60.85 & 6 & \textbf{68.90} & \textbf{1} & \underline{66.74} & \underline{2} \\
\texttt{cGv19pUNet\tiny{1-1}} \hspace{-0.2cm} & 84.15 & 4 & \underline{69.21} & \underline{2} & \textit{68.94} & \textit{3} & 62.27 & 4 & 67.12 & 5 & \textbf{66.86} & \textbf{1} \\ \hline
\texttt{MOvpUNet} \hspace{-0.2cm} & 85.53 & $\star$ & 70.17 & $\star$ & 70.39 & $\star$ & 64.77 & $\star$ & 69.92 & $\star$ & 68.78 & $\star$ \\ \hline
\end{tabular}
\caption{Scoreboard and ranking of \texttt{DeepMedic} \cite{kamnitsas2017efficient}, \texttt{denseVNet} \cite{gibson2018automatic}, \texttt{UNet} \cite{ronneberger2015unet}, \texttt{v16UNet}, \texttt{v16pUNet} \cite{conze19isbi}, \texttt{v19UNet}, \texttt{v19pUNet}, \texttt{v19UNet$+$} \cite{zhou2018unet}, \texttt{v19pUNet$+$} \cite{zhou2018unet}, \texttt{UNet\tiny{1-1}} \cite{yan19cascaded}\textcolor{black}{,} proposed \texttt{v16pUNet\tiny{1-1}},  \texttt{v19pUNet\tiny{1-1}}, \texttt{cGv16pUNet\tiny{1-1}}, \texttt{cGv19pUNet\tiny{1-1}} and \texttt{MOvpUNet} for healthy abdominal organ (liver, \textcolor{black}{right kidney}, \textcolor{black}{left kidney}, spleen) segmentation in CT and MR images. Bold, underline and italic results indicate first, second and third best scores.} \vspace{-0.35cm}
\label{tab::tab-3}
\end{table*}

As visually shown in Fig.\ref{fig::fig-8}, many anomalies are present using standard networks with over (under-) detection issues for \texttt{DeepMedic} (\texttt{denseVNet}) in T2-SPIR. \texttt{v19pUNet\tiny{1-1}} and \texttt{cGv19pUNet\tiny{1-1}} show a better behavior than \texttt{v19pUNet} and \texttt{v19pUNet$+$} in accurately fitting organ \textcolor{black}{extents} and offering plausible shape consistency, especially for bases and apexes where organs appear smaller. Despite visually similar performance compared to \texttt{v19pUNet\tiny{1-1}}, \texttt{cGv19pUNet\tiny{1-1}} appears slightly better in providing realistic organ \textcolor{black}{contours}.

\noindent \textbf{Towards better multi-organ segmentation.} Under the team name \texttt{PDKIA}, the proposed \texttt{cGv19pUNet\tiny{1-1}} pipeline enabled us to win three CHAOS competition categories: liver CT, liver MR and multi-organ MR segmentation \cite{kavur2020chaos}. Nevertheless, since global findings are verified with varying degrees depending on the concerned modality or organ, we provide in Tab.\ref{tab::tab-3} an overall evaluation through scores and rankings for CT liver, MR liver, right kidney, left kidney, spleen as well as multi-organ segmentation. \texttt{cGv19pUNet\tiny{1-1}} indeed appears as the best strategy for MR multi-organ delineation purposes that reinforces the idea that combining deeper (\texttt{v19}) cascaded partially pre-trained convolutional and adversarial networks globally strengthens the generalization abilities of deep learning pipelines. Except for left kidney \textcolor{black}{where} \texttt{v16pUNet} performs the best \textcolor{black}{($64.05\%$)}, the first rank is always attributed to one of the proposed cascaded pre-trained scheme: \texttt{v16pUNet\tiny{1-1}} for CT liver, \texttt{v19pUNet\tiny{1-1}} for MR liver and right kidney \textcolor{black}{($69.45\%$)}, \texttt{cGv16pUNet\tiny{1-1}} for MR spleen \textcolor{black}{($68.9\%$)} and \texttt{cGv19pUNet\tiny{1-1}} for MR multi-organ \textcolor{black}{($66.86\%$)} segmentation. 

\begin{figure*}
\scriptsize
\hspace{-0.2cm} \begin{minipage}{0.5\linewidth}
\begin{tabular}{cccccc}
\rotatebox{90}{\textcolor{white}{------------} CT} &
\hspace{-0.3cm} \includegraphics[height=2.63cm]{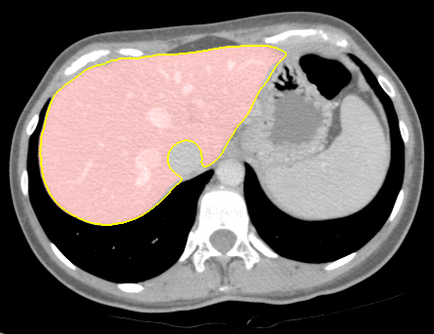} &
\hspace{-0.4cm} \includegraphics[height=2.63cm]{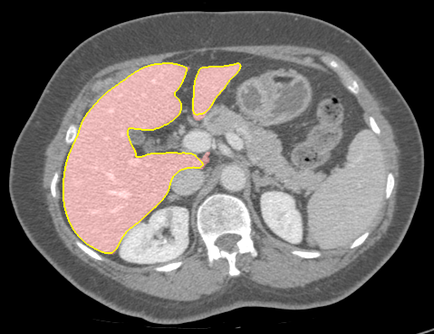} &
\hspace{-0.4cm} \includegraphics[height=2.63cm]{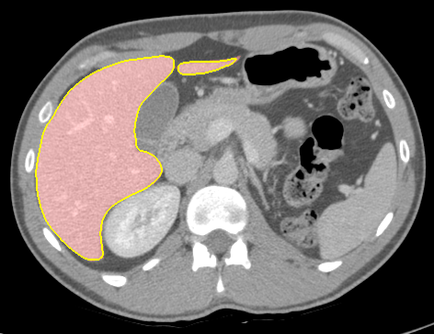} &
\hspace{-0.4cm} \includegraphics[height=2.63cm]{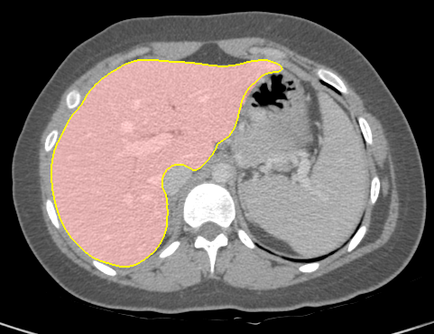} &
\hspace{-0.4cm} \includegraphics[height=2.63cm]{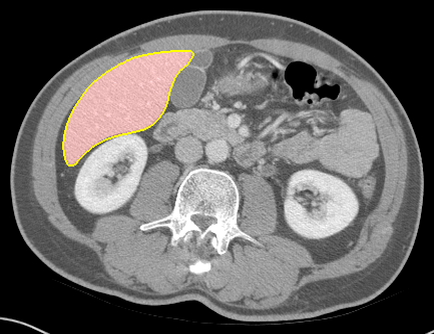} \cr
\rotatebox{90}{\textcolor{white}{------} T1-DUAL\texttt{in}} &
\hspace{-0.3cm} \includegraphics[height=2.63cm]{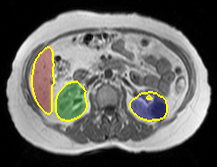} &
\hspace{-0.4cm} \includegraphics[height=2.63cm]{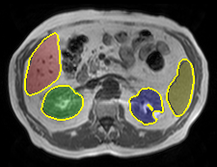} &
\hspace{-0.4cm} \includegraphics[height=2.63cm]{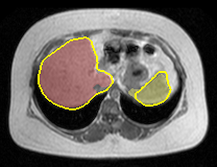} &
\hspace{-0.4cm} \includegraphics[height=2.63cm]{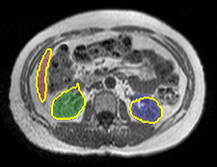} &
\hspace{-0.4cm} \includegraphics[height=2.63cm]{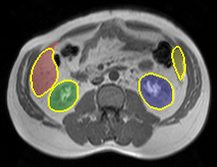} \cr
\rotatebox{90}{\textcolor{white}{-----} T1-DUAL\texttt{out}} &
\hspace{-0.3cm} \includegraphics[height=2.63cm]{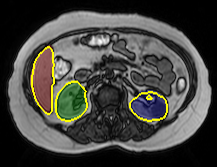} &
\hspace{-0.4cm} \includegraphics[height=2.63cm]{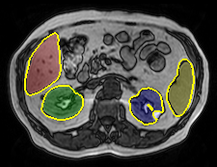} &
\hspace{-0.4cm} \includegraphics[height=2.63cm]{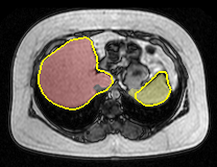} &
\hspace{-0.4cm} \includegraphics[height=2.63cm]{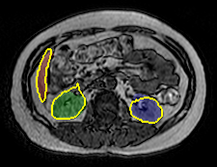} &
\hspace{-0.4cm} \includegraphics[height=2.63cm]{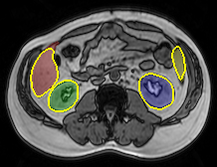} \cr
\rotatebox{90}{\textcolor{white}{----------} T2-SPIR} &
\hspace{-0.3cm} \includegraphics[height=2.63cm]{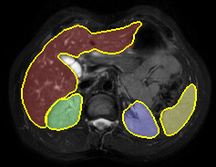} &
\hspace{-0.4cm} \includegraphics[height=2.63cm]{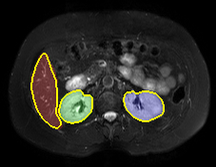} &
\hspace{-0.4cm} \includegraphics[height=2.63cm]{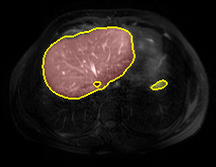} &
\hspace{-0.4cm} \includegraphics[height=2.63cm]{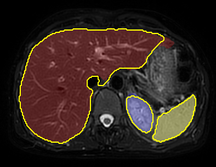} &
\hspace{-0.4cm} \includegraphics[height=2.63cm]{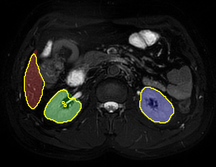} \cr
\end{tabular}
\end{minipage} \vspace{-0.1cm} \\
\begin{center}
\textcolor{color5}{-----} \texttt{MOvpUNet}
\textcolor{color14}{$\blacksquare$} liver
\textcolor{color11}{$\blacksquare$} right kidney
\textcolor{color12}{$\blacksquare$} left kidney
\textcolor{color13}{$\blacksquare$} spleen
\end{center} \vspace{-0.1cm}
\caption{Liver CT and abdominal multi-organ MRI (T1-DUAL\texttt{in/out}, T2-SPIR) segmentation using the proposed \texttt{MOvpUNet}. Liver, right kidney, left kidney and spleen groundtruth delineations are superimposed in red, green, blue and yelow colors.} \vspace{-0.3cm}
\label{fig::fig-9}
\end{figure*}

By combining the best sequence- and organ-specific networks towards better Multi-Organ (\texttt{MO}) MR segmentation, we obtain the so-called \texttt{MOvpUnet} computational model including \texttt{v16pUNet\tiny{1-1}} for liver in CT (Tab.\ref{tab::tab-1}) as well as respectively for T1-DUAL and T2-SPIR \texttt{cGv19pUNet\tiny{1-1}} and \texttt{v19pUNet\tiny{1-1}} for liver, \texttt{v19pUNet} and \texttt{v19pUNet\tiny{1-1}} for right kidney, \texttt{v16pUNet} and \texttt{cGv16pUNet\tiny{1-1}} for left kidney, \texttt{v19pUNet$+$} and \texttt{cGv16pUNet\tiny{1-1}} for spleen (Tab.\ref{tab::tab-2}). The global ranking \textcolor{black}{score} reached by \texttt{cGv19pUNet\tiny{1-1}} \textcolor{black}{for} multi-organ MR segmentation is further improved about $2\%$ with \texttt{MOvpUnet}, up to $68.78\%$ (Tab.\ref{tab::tab-3}). Visually comparing manual and \texttt{MOvpUnet} delineations in Fig.\ref{fig::fig-9} further supports the validity of our combined computational model. Outstanding performance is reached in terms of boundary adherence and shape consistency which suggests that integrating \texttt{MOvpUnet} as \textcolor{black}{a} guidance tool into clinical routine could greatly help clinicians for abdominal image interpretation. \vspace{-0.2cm}

\section{Conclusion}
\label{sec:sec6}

This work tackles fully-automated abdominal organ CT and MR segmentation with deep learning. Standard segmentation networks are extended to cascades of partially pre-trained deep convolutional encoder-decoders. Encoder fine-tuning from \textcolor{black}{a large amount} of non-medical images improves predictive performance while alleviating data scarcity limitations. The cascaded architecture exploits multi-level contextual information through auto-context and end-to-end training. Such model is used as generator in a conditional generative adversarial network to further encourage the generative part to provide plausible organ delineations. Results highlight promising performance by outperforming state-of-the-art encoder-decoder schemes. Employed for the Combined Healthy Abdominal Organ Segmentation (CHAOS) challenge, our contributions reached the first rank for liver CT, liver MR and multi-organ MR segmentation competition categories. The proposed pipeline \textcolor{black}{has} the potential to support guidance for abdominal image interpretation, clinical decision making and patient care improvement while avoiding manual delineation efforts. Further work includes the evaluation of such deep models to other anatomical structures from the abdomen (pancreas, gallbladder) and the gastro-intestinal tract (esophagus, stomach, duodenum) arising from healthy \textcolor{black}{or} pathological subjects. More globally, our pipeline could be easily extended to other tissue types and imaging modalities to provide relevant clinical decision support. Methodological perspectives on unpaired cross-modality medical image segmentation with compact architectures could deserve further investigation to take advantage of multi-tasking properties of deep models as well as \textcolor{black}{a} larger amount of available data. Extending adversarial frameworks to incorporate anatomical priors \textit{via} topological or shape \textcolor{black}{constraints} should also offer new insights to manage the strong diversity of abdominal organ appearance.

\ifCLASSOPTIONcaptionsoff
  \newpage
\fi

\bibliographystyle{IEEEtran} \vspace{-0.05cm}
\bibliography{conze-TBME-2019-chaos}

\end{document}